\documentclass{aastex631}
\usepackage[T1]{fontenc}

\usepackage{color}
\usepackage{subfigure}
\usepackage{CJK}

\received{March 8, 2023}
\revised{Oct 12, 2023}
\accepted{Oct 12, 2023}

\submitjournal{ApJ}

\shorttitle{Two Different Evolutionary Patterns}
\shortauthors{Haerken H et al.}
\graphicspath{{./}{figures/}}

\begin{document}
\begin{CJK*}{UTF8}{gbsn}
\title{Discovery of Two Different Full Disk Evolutionary Patterns of M-type T Tauri Stars with LAMOST DR8}

\correspondingauthor{Guang-Wei Li}
\email{lgw@bao.ac.cn}
\correspondingauthor{Fuqing Duan}
\email{fqduan@bnu.edu.cn}

  \author[0000-0001-7513-3316]{Hasitieer Haerken (哈斯铁尔$\cdot$哈尔肯)}
  \affiliation{ School of Artificial Intelligence of Beijing Normal University, 
	No.19, Xinjiekouwai St, Haidian District, Beijing, 100875, China}
  \affiliation{Key Laboratory of Space Astronomy and Technology, National Astronomical Observatories, Chinese Academy of Sciences, 20A Datun Road, Chaoyang District, Beijing, 100101, China}

\author[0000-0001-7515-6307]{Guang-Wei Li (李广伟)}
\affiliation{Key Laboratory of Space Astronomy and Technology, National Astronomical Observatories, Chinese Academy of Sciences, 20A Datun Road, Chaoyang District, Beijing, 100101, China}

  \author[0000-0002-4265-8834]{Min Li (李敏)}
  \affiliation{ School of Information Technology and Engineering, Tianjin University of Technology and Education, 1310 Dagu South Road, Hexi District, Tianjin, 300222, China}

  \author[0000-0002-3849-8532]{Fuqing Duan (段福庆)}
  \affiliation{ School of Artificial Intelligence of Beijing Normal University, 
	No.19, Xinjiekouwai St, Haidian District, Beijing, 100875, China}

 \author{Yongheng Zhao (赵永恒)}
\affiliation{National Astronomical Observatories, Chinese Academy of Sciences, 20A Datun Road, Chaoyang District, Beijing, 100101, China}

\begin{abstract} 
   The full disk, full of gas and dust, determines the upper limit of planet masses, and its lifetime is critical for planet formation, especially for giant planets. In this work, we studied the evolutionary timescales of the full disks of T Tauri stars (TTSs) and their relations to accretion. Combined with Gaia EDR3, Two Micron All Sky Survey, and Wide-field Infrared Survey Explorer data, 1077 disk-bearing TTS candidates were found in LAMOST DR8, and stellar parameters were obtained. Among them, 783 are newly classified by spectra as classical T Tauri stars (CTTSs; 169) or weak-lined T Tauri stars (WTTSs). Based on EW and FWHM of H$\alpha$, 157 TTSs in accretion were identified, with $\sim$ 82\% also having full disks. For TTSs with $M < 0.35 M_{\sun}$, about 80\% seem to already lose their full disks at $\sim$ 0.1 Myr, which may explain their lower mass, while the remaining 20\% with full disks evolve at similar rates of non-full disks within 5 Myr, possibly suffice to form giant planets. The fraction of accreting TTSs to disk-bearing TTSs is stable at $\sim$10\% and can last $\sim$ 5-10 Myr, suggesting that full disks and accretion evolve with similar rates as non-full disks. For TTSs with $M > 0.35 M_{\sun}$, almost all full disks can survive more than 0.1 Myr, most for 1 Myr and some even for 20 Myr, which implies planets are more likely to be formed in their disks than those of $M < 0.35 M_{\sun}$, and thus M dwarfs with $M > 0.35 M_{\sun}$ can have more planets. The fraction of full-disk TTSs to disk-bearing TTSs decreases with age following the relation $f\propto t^{-0.35}$, and similar relations existed in the fraction of accreting TTSs and the fraction of full-disk CTTSs, suggesting faster full disks and accretion evolution than non-full disks. For full disk stars, the ratio of accretion of lower-mass stars is systematically lower than that of higher-mass stars, confirming the dependence of accretion on stellar mass, which may be reflective of an observational bias in the detection of accretion levels, with the lower-mass stars crossing below the detection threshold earlier than higher-mass stars.
\end{abstract}


\section{Introduction} 
\label{sec: intro}

The evolution of circumstellar disks is essential for the formation of stars and their planets. The disk accretion dominates in the early stage \citep{2014prpl.conf..475A} and the disk dissipates over time for several reasons, including photoevaporation, planet formation, grain growth, etc \citep{2006ApJ...639..275K,2014prpl.conf..475A, 2021A&A...655A..18G}. According to mid-infrared photometric observations, the evolutionary process of disks can be roughly classified into four main stages: the full disk at the early stage \citep{2012ApJ...758...31L}, transitional disk with inner holes and gaps formed by dust condensation and dynamic removal of newly formed planets \citep{2019A&A...629A..67S}, evolved disks that are becoming optically thin \citep{2010ApJS..186..111L, 2012ApJ...758...31L}, and debris disks, the last stage in the disk evolution, which is maintained by collisional cascades \citep{2012ApJ...751L..17M}. In this work, we refer to the transitional, evolved, and debris disks as non-full disks.
\par
In circumstellar disks, planet formation starts with small dust, which slowly grows into pebbles, then condenses into planetesimals, and eventually forms larger planetary embryos through collisions \citep{2010A&A...513A..79B, 2013ApJ...764..146G, 2016ApJ...831....8K}. When the mass of the planet's core is large enough, it will capture the gas from the accretion disk, and a planet is formed \citep{2012AREPS..40..251M}. The time required for planet formation varies from model to model, e.g. \citet{1996Icar..124...62P} suggested that it takes 8 Myr to form a Jupiter, while \citet{2000ApJ...537.1013I} thought it would take as long as 100 Myr. Some studies suggested that it takes several megayears to form planets larger than 10 M$_{\oplus}$ by core accretion \citep{2004AJ....128..805S, 2005Icar..179..415H, 2012ApJ...745...19K, 2022ASSL..466....3R}. The key stages of planet formation are thought to be between 1 and 10 Myr \citep{2019MNRAS.486.4590C} and can even be traced back to the earliest few 0.1 Myr according to the meteorite record 
\citep{2022ASSL..466....3R}. In the age range of 3-5 Myr, pebbles are observed more frequently in disks, probably because disk cleaning leaves pebbles and planetesimals behind \citep{2017AJ....153..240A, 2019A&A...627A..83L}. PDS 70b is the first directly imaged newly forming gas giant exoplanet \citep{2018A&A...617L...2M} with a T Tauri star (TTS) as its host star \citep{2006A&A...458..317R, 2018ApJ...863L...8W}, and several planetary candidates are being successively discovered in accretion disks of TTSs \citep{2018A&A...619A.171B, 2018ApJ...869L..50P, 2020ApJ...892..111F,2022SCPMA..6569511L}. 
\par
The lifetime of disks of the TTSs appears to be strongly correlated with the stellar mass, as disks around more massive stars tend to dissipate more rapidly than those around less massive stars \citep[e. g.][]{2018ApJ...869...72Y,2022AJ....163...25L,2022ApJ...939L..10P}. Most K- and M-type stars are thought to be surrounded by disks at about 1 Myr \citep{2016ApJ...827..142B}. The e-folding timescale of the disks for K-type TTSs is estimated to be 4.7 Myr \citep{2016MNRAS.461..794P}, while for M-type stars the disk should last much longer \citep{2020ApJ...890..106S}, and the disk fraction is even still as high as about 9\% at 20 Myr. \citet{2022ApJ...939L..10P} suggested that low-mass stars have a median disk lifetime of 5-10 Myr. However, we should be aware that all the timescales mentioned above are for disks,  not just for full disks. The timescale of full disks should be shorter.
\par
The occurrence of Earth-like and super-Earth planets around M-type stars is high \citep{2021ApJ...920L...1M}, while giant planets are rare \citep{2022A&A...664A.180S}. \citet{2014ApJ...781...28M} estimated that about 6.5\% $\pm$ 3.0\% of M dwarfs have a giant planet with a mass between 1 and 13 $M_{\rm J}$. Many models suggest a minimum host mass of 0.3 $M_{\sun}$ for giant planet formation \citep{2019A&A...632A...7L,2022A&A...664A.180S}, and for stars with $M < 0.25 M_{\sun}$, even with a sufficiently massive disk it is not possible to form any planets with masses above 5 $M_{\oplus}$ \citep{2020MNRAS.491.1998M, 2022A&A...664A.180S}. However, the recent discovery of sub-Neptunian in the M4 type \citep{2021A&A...653A..97W} and giant planets in the M5 type \citep{2022A&A...663A..48Q} poses a challenge to theoretical models of planet formation. The giant planets should be formed in full disks, so to explore why giant planets are rare around low-mass stars, the timescale of their full disks should be known.
\par
By using low-resolution spectra (LRS) of the Large Area Multi-Objective Fiber Optic Spectroscopic Telescope (LAMOST), we will try to find the M-type TTSs undergoing accretion which is a signal of early-stage disks, and explore the timescale of the full disks around M-type TTSs with Wide-field Infrared Survey Explorer (WISE) and Two Micron All Sky Survey (2MASS) data. This paper is organized as follows. In Section \ref{sec: sec2}, we introduce the data used in this paper. The method to search for reliable TTS candidates is given in Section \ref{sec: sec3}. In Section \ref{sec: sec4}, we analyze the full disk and accretion of TTSs and study the relationship between them. Stellar atmospheric parameters obtained from LAMOST LRS are given in Section \ref{sec: sec5}. In Section \ref{sec: sec6} we analyze the evolution of the full disk of the M-type TTSs. We presente a discussion in Section \ref{sec: sec7} and summarize our work in Section \ref{sec: sec8}.

\section{Data} \label{sec: sec2}	
  \begin{figure*}[ht!]
    \plotone{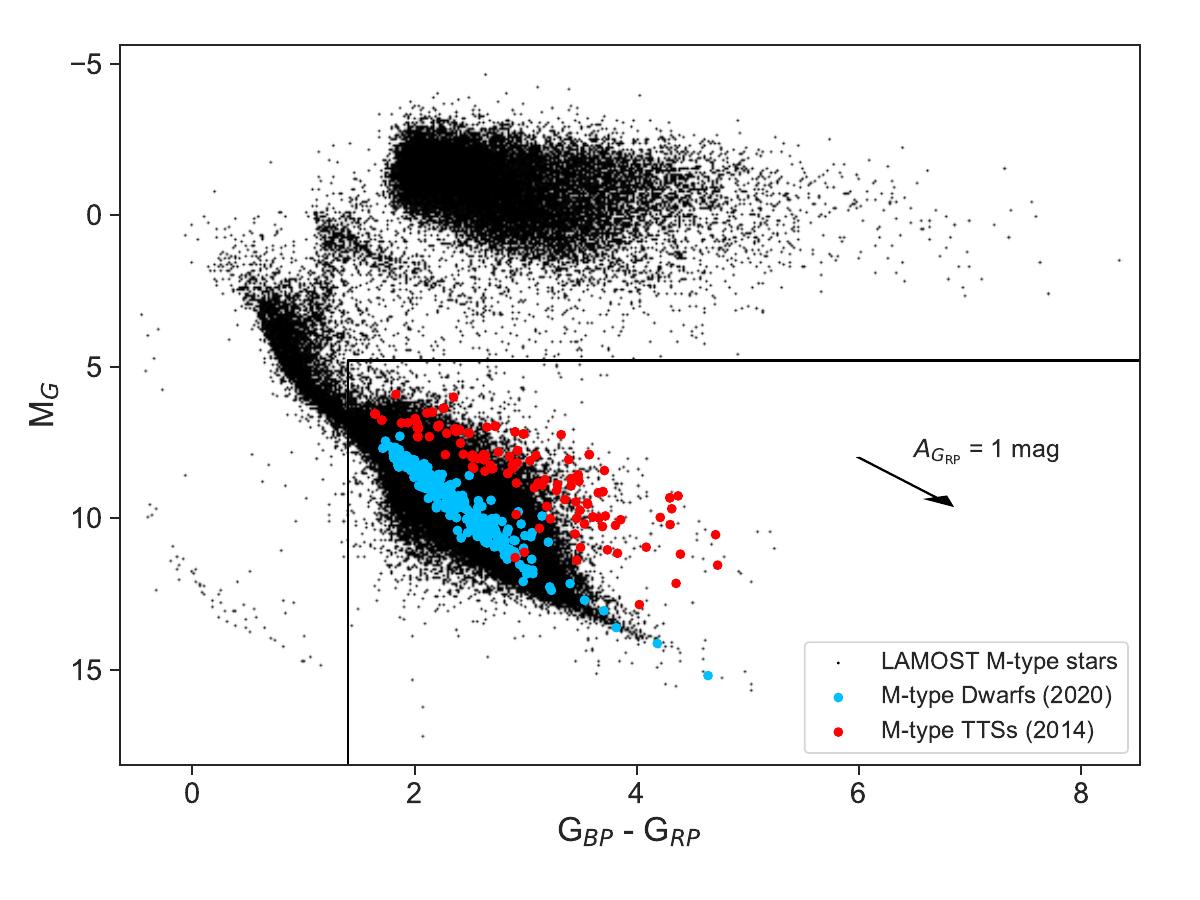}
    \caption{The diagram of G$_{\rm BP}$-G$_{\rm RP}$ vs. M$_{\rm G}$. The stars from the LAMOST DR8 M-type catalog, M-type dwarfs from 
    \citep{2020A&A...642A.115C} and M-type TTSs from \citet{2014ApJ...786...97H} are shown in black, blue, and red, respectively.
      \label{fig: LAMOST data}}
  \end{figure*} 
  
\subsection{The LAMOST spectra}

LAMOST \citep{2012RAA....12..735D,2012RAA....12..723Z, 2014IAUS..298..310L, 2015MNRAS.448..855Y, 2017MNRAS.467.1890X}, combining a large aperture with a wide field of view, has 4000 fibers placed on its focal plane to obtain the spectra of 4000 objects simultaneously, providing massive amounts of stellar spectra for scientific research. LAMOST conducts regular surveys, without objects of interest, aiming to study the structure of Galactic halo and disk components, covering a significant volume of some star-forming regions.
\par
In LAMOST DR8 \footnote{\url{http://www.lamost.org/dr8/v1.0/}}, there are more than 10 million LRS ($R \sim 1800$), with an M-type star catalog containing about 770,000 spectra of M-type stars. The wavelength range is from 3690-9100 \AA \ and covers the Balmer and Ca II emission lines, as well as the Li I, Na I, and K I absorption lines for the spectral identification of TTSs. 
\par
LAMOST also collects medium-resolution spectra (MRS; $R\sim7500$), aiming to study such as star formation, host stars of exoplanets, emission nebulae, and open clusters, covering many star-forming regions such as Taurus, Orion, Perseus, and some associations \citep{2020arXiv200507210L}. Therefore, the LAMOST MRS is also used in this work, if available.
\par

\subsection{Photometric Data} \label{subsec: Photometric data}
The photometric data we used are $G$, $G_{\rm BP}$, and $G_{\rm RP}$ from Gaia EDR3 \citep{2020yCat.1350....0G}, $J$, $H$, and $Ks$ from 2MASS \citep{2003yCat.2246....0C,2006AJ....131.1163S} and W1, W2, W3 and W4 from ALLWISE \citep{2014yCat.2328....0C}.\par
We firstly cross-matched the LAMOST M-type star catalog with Gaia EDR3 with a radius of 3" and obtained the absolute magnitude of $G$: $M_G = G - 5\lg (1000/\varpi) +5$, where $\varpi$ is the parallax given by Gaia EDR3 in milliarcsecond. The color-magnitude diagram (CMD) of these LAMOST M-type stars is shown in Figure \ref{fig: LAMOST data} in black, and some known dwarfs and TTSs are overplotted in blue and red as guides, respectively. The M dwarfs are from the CARMENES input catalog \citep{2020A&A...642A.115C}, which contains 2210 nearby M-type stars, but the data marked as ``Multiple,'' ``Young,'' and ``Excess'' in the catalog were removed. TTSs are from the catalog in \citet{2014ApJ...786...97H}, which contains 281 well-studied near-neighbor TTSs, but the sources with mismatched photometric data, double or multiple stars, and the stars with spectral types earlier than M0 were removed. From Figure \ref{fig: LAMOST data}, we can see that M-type dwarfs and M-type TTSs are located in the region of $G_{\rm BP}-G_{\rm RP} > 1.4$ and $M_G > 4.8$. Now, we require our samples should satisfy all the following conditions:
\begin{itemize}
\item If two or more stars in Gaia EDR3 and ALLWISE match the same LAMOST M-type star in the radius of 3", only the brightest one is adopted because most of the light recorded by LAMOST should be from the brightest star. 
\item $G$, $G_{\rm BP}$, $G_{\rm RP}$, proper motion, parallax $\varpi$, and parallax error $\sigma_{\varpi}$ should be available and satisfy $\varpi / \sigma_{\varpi} > 5$.
 \item $G_{\rm BP}-G_{\rm RP} > 1.4$, and $M_G>4.8$. 
 \item In the ALLWISE catalog, the $ccf$ flag of the source should be ``00XX,'' which means W1 and W2 data are not affected by known artifacts, and the $qph$ flag should be ``AAXX,'' which means the signal-to-noise ratios (S/Ns) of W1 and W2 are greater than 10.
\item $J$, $H$, and $Ks$ are available in the ALLWISE catalog.
\end{itemize}
As a result, 506,735 stars were obtained from the M-type star catalog of LAMOST DR8.
\par
We also require these stars should be in star-forming regions, which are constrained in the area of 140\degr $\leq l \leq$ 220\degr \ and -40\degr $\leq b \leq$ 10\degr \ because most TTSs are still located in star-forming regions, though some are discovered outside the star-forming regions \citep{2020ApJ...900...14F, 2020MNRAS.495.3104S,2020ApJS..251...18T}. Finally, 96,541 LAMOST stars were retained.

\section{To find TTS candidates} \label{sec: sec3}

  \begin{figure*}[ht!]
  \centering
    \includegraphics[width=1\textwidth,trim=0 0 0 0,clip]{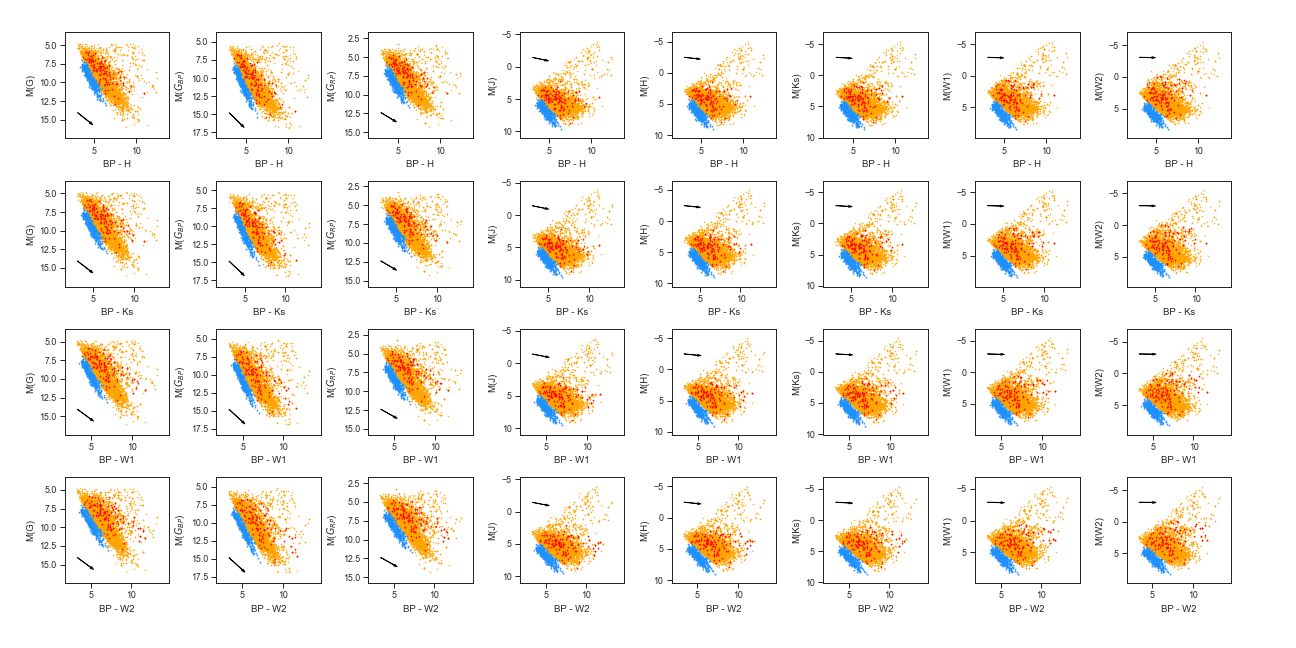}
    \caption{The CMDs of TTS candidates obtained by the triplet network algorithm. In each panel, M dwarfs and TTSs used to train the algorithm are shown by blue and red triangles, respectively, and orange dots are TTS candidates found by the algorithm. The black arrows are the reddening vectors with 
    $A_{G_{\rm RP}}$ = 1 mag.
      \label{fig: TTS_CMD}}
  \end{figure*} 

\subsection{The Triplet Network Algorithm} 
The triplet network model \citep{2014arXiv1412.6622H} is adopted to separate LAMOST TTSs from dwarfs. It works by mapping samples to the embedding space through a learning function, modeling the similarity between samples, and completing the classification task by maximizing the inter-class distance and minimizing the intra-class distance. 
\par
The features used in the algorithm are the absolute magnitudes of $G$, $G_{\rm BP}$, $G_{\rm RP}$, $J$, $H$, $Ks$, W1, and W2, and 28 colors from the combinations of these magnitudes. We tried to reduce the dimensionality of the data, but it did not help much, which is consistent with \citet{2019MNRAS.487.2522M}. The stars used to train the algorithm are from the M dwarfs (blue dots) and M-type TTSs (red dots) in Figure \ref{fig: LAMOST data}. To distinguish well between TTSs and dwarfs, we only use those that can be well separated in their CMDs and obtained 113 TTSs and 444 dwarfs, respectively, which are shown in blue and red in Figure \ref{fig: TTS_CMD}. Since our data are widely distributed in the sky and it is difficult to obtain their true extinction values, the algorithm aims to select out TTSs with significant color excesses and higher luminosities than dwarfs even in the presence of heavy extinction. In each panel of Figure \ref{fig: TTS_CMD}, the reddening vector is marked by the black arrow with $A_{G_{\rm RP}} = 1$ mag, using the relative extinction relations from \citet{2019ApJ...877..116W}. 
\par
The framework of the fully connected triplet network is similar to \citet{2014arXiv1412.6622H}, which consists of three identical sub-networks with shared weights. The input is three samples $\left(X_p, X_m, X_n \right)$, where X denotes 36-dimensional data of a star consisting of absolute magnitudes and colors, and the output of the network is the label of the category that the star belongs to. Each sub-network has three dense layers with 36, 36, and 18 nodes, respectively, which were obtained after several attempts. A ReLu activation function follows each dense layer and a dropout layer with a dropout rate equal to 0.1 is added after each of the first two dense layers to prevent overfitting. The network inputs the learned features into the distance layer, which calculates the distance between two samples in the embedding space as follows:
  \begin{equation}
    D_{\rm w} \left ( X_{\rm 1}, X_{\rm 2}\right ) =\left \| D_{\rm w} \left ( X_{\rm 1}\right )-D_{\rm w} \left (X_{\rm 2}\right ) \right \|_{\rm 2},
  \end{equation}
where $X_{\rm 1}$ and $X_{\rm 2}$ are the embedding features of the sample pairs. Then the distances $D_{\rm w} \left ( X_{\rm m}, X_{\rm p}\right )$ and $D_{\rm w} \left ( X_{\rm m}, X_{\rm n}\right )$ will be sent to the comparator and the triplet loss function used is
\begin{equation}
    L\left ( W, X_{\rm p}, X_{\rm m}, X_{\rm n}, Y \right ) \!=\! L_{\rm P} \!+\! L_{\rm N},
  \end{equation}
  \begin{equation}
    L_{\rm P} = \frac{1}{2}\!\times\! Y \!\times\! D_{\rm W}\left ( X_{\rm p}, X_{\rm m}\right ) ^2,
  \end{equation}
  \begin{equation}
    L_{\rm N} \!=\! \frac{1}{2}\! \times \! \left ( 1\!-\!Y \right ) \!\times\! D_{\rm N}\left ( X_{\rm m}, X_{\rm n}\right ) ^2,
  \end{equation}
  \begin{equation}
    D_{\rm N}\left ( X_{\rm m}, X_{\rm n}\right ) \!=\! max\left ( 0, margin\!-\!D_{\rm W}\left (X_{\rm m}, X_{\rm n} \right ) \right ),
  \end{equation}
where $Y$ represents the label of the sample, and the $margin$ value is set 1. 
\par
The input sample is constructed by randomly selecting a sample $X_m$, a positive sample $X_p$ from 113 TTSs, and a negative sample $X_n$ from 444 dwarfs, with $Y = 1$ if $X_m$ is from the same category as $X_p$, otherwise $Y = 0$. We divided the created samples into training and validation sets in the ratio of 7:3 and determined the settings of each parameter experimentally. The optimizer used is RMSpropr \citep{2016arXiv160904747R} and the parameter settings are batch = 128 and epochs = 50. The network converges when the epochs are equal to 10. In the validation set, the accuracy, precision, recall rate, and F1 score are all higher than 99.9\%, which is mainly due to clean samples, valid features, and suitable algorithms, rather than overfitting.
\par
The test data are from 96,541 samples discussed in Section \ref{subsec: Photometric data}. For each test sample $X_{\rm test}$, we randomly choose 50 TTSs ($X_p$) and 50 dwarfs ($X_n$) to form 50 input samples ($X_p$, $X_{\rm test}$, $X_n$) and feed them into the network. Each run of the network yields 50 labels for each $X_{\rm test}$. To minimize misclassification due to randomness in network and sample selection, we trained the neural network using three different sets of training samples to get 50 labels for each test run, so that a total of 150 labels can be obtained for each test sample. By voting, a star was accepted as a TTS candidate only if it was labeled as ``TTS'' in at least 85\% of the 150 labels. Finally, 13,404 stars are obtained and shown by orange dots in Figure \ref{fig: TTS_CMD}.

\subsection{Identification of TTSs} \label{subsec: Membership}
EW(H$\alpha$) is the most well-used indicator in TTS spectral analysis. EW(H$\alpha$) $\geq 5$ \AA \ is firstly given by \citet{1962AdA&A...1...47H} to distinguish between dMe stars and TTSs, and later the threshold of EW(H$\alpha$) = 10 \AA \ was used to distinguish between classical T Tauri stars (CTTSs) and Weak-lined T Tauri stars (WTTSs). Then \citet{2009AA...504..461F} gave finer criteria based on EW(H$\alpha$) of different spectral types to separate CTTSs from TTSs, applicable to LRS. 
\par
We visually examined the spectral data of all TTS candidates and found that most of the data with H$\alpha$ line S/N greater than 3 have EW(H$\alpha$) $\geq$ 5 \AA. Therefore, we set a threshold of EW(H$\alpha$) $\geq$ 5 \AA \ to remove the contamination of dMe stars and also to guarantee the H$\alpha$ line S/N. 
\par
In the area of 140\degr $\leq l \leq$ 220\degr \ and -40\degr $\leq b \leq$ 10\degr, there are eight big star-forming regions: Taurus, Perseus, Orion complex, L1616, California, Monoceros OB1 (Mon OB1), MBM molecular cloud, and Cassiopeia-Taurus OB association (Cas-Tau). The potential association between the TTS candidates and the star-forming regions is determined from the galactic coordinates and the distances calculated from Gaia parallaxes, where the locations and distances of the star-forming regions are taken from \citet{1999AJ....117..354D}, \citet{2014ApJ...786...29S}, and \citet{2019ApJ...879..125Z, 2020A&A...633A..51Z}:
\begin{itemize}
	\setlength{\itemsep}{0pt}
	\setlength{\parsep}{0pt}
	\setlength{\parskip}{0pt}
    \item Taurus is located in $l \sim$ (166\degr, 176\degr) and $b \sim$ (-19\degr, -13\degr) with a distance around 140 pc;
    \item Perseus is located in $l \sim$ (157.\degr5, 161.\degr5) and $b \sim$ (-22\degr, -16\degr) with distance of $\sim$ 290 pc;
    \item The Orion complex is one of the largest nearby star-forming regions, located in $l \sim$ (191.\degr5, 215\degr) and $b \sim$ (-20.\degr5, -7\degr) with a distance of $\sim$ 400 pc;
    \item L1616, an on-going star-forming region, is centered at $l \sim$ 203.\degr5 and $b \sim$ -24.\degr8, with a distance of $\sim$ 390 pc;
    \item California locates in $l \sim$ (160\degr, 166\degr) and $b \sim$ (-10\degr, -7.\degr5) and the distance is $\sim$ 470 pc;
    \item Mon OB1 is centered at $l \sim$ (199\degr, 204\degr) and $b \sim$ (0\degr, 3\degr) and the distance is $\sim$ 745 pc, with NGC 2264 as the core;
    \item The MBM molecular cloud contains lots of small groups with relatively dispersed distributions, so specific positional and distance information is not listed here and details can be found in \citet{2014ApJ...786...29S}, \citet{2019ApJ...879..125Z}, and \citet{2021ApJS..256...46S};
    \item Cas-Tau covers a large area on the celestial plane and has a wide range of distances from 125-300 pc \citep{1999AJ....117..354D, 2020ApJS..251...18T}.
\end{itemize}

 \begin{figure*}[t]
	\plotone{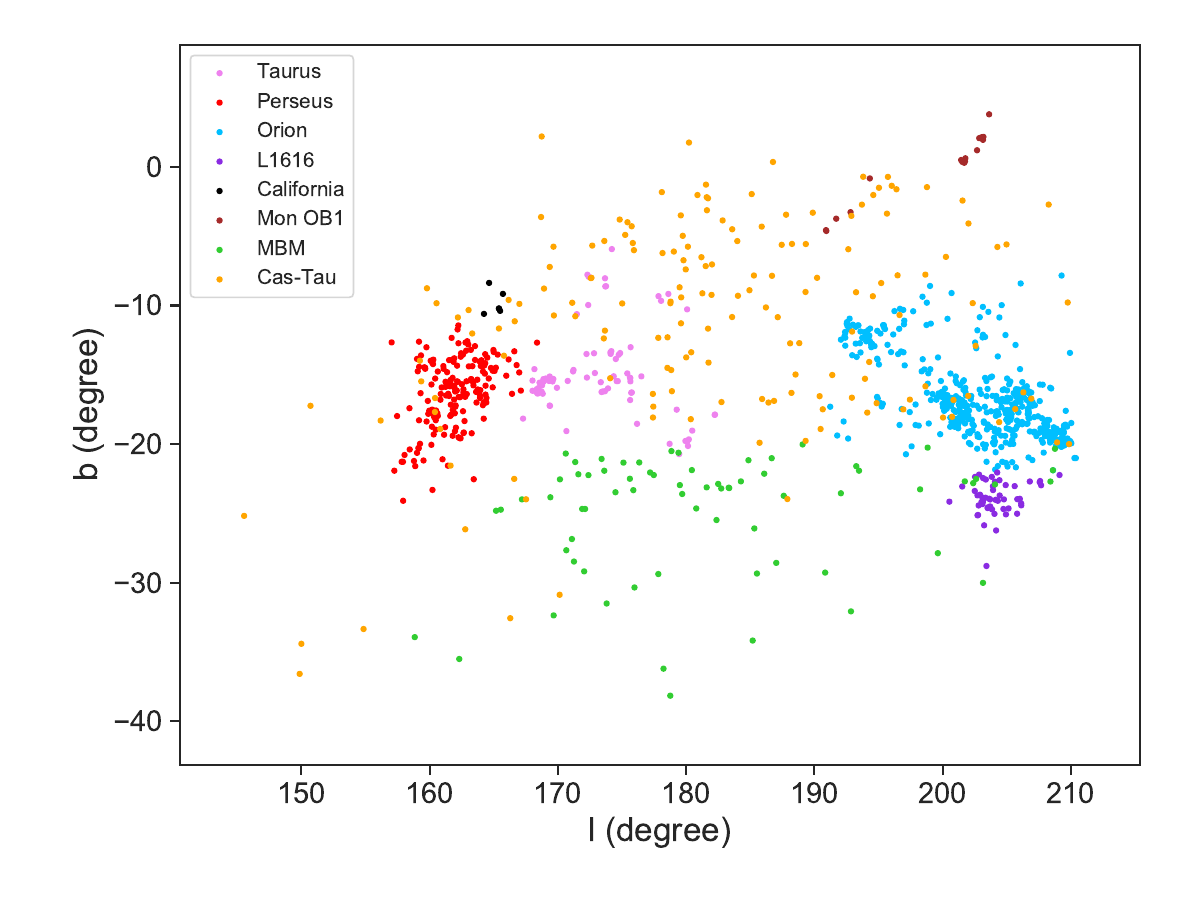}
	\caption{Spatial distribution of 1077 TTSs.} \label{fig: sky_distribution}
\end{figure*} 
\par
Restricting the TTS candidates to be located in star-forming regions, we finally obtained 1077 reliable TTS candidates (hereafter referred to as TTSs) as shown in Figure \ref{fig: sky_distribution}, of which 78 in Taurus, 187 in Perseus, 500 in Orion complex, 53 in L1616, six in California, 17 in Mon OB1, 70 in MBM molecular cloud, and 166 in Cas-Tau. 
\par
We visually inspected their spectral types given by the LAMOST DR8, and for the stars with obviously wrong spectral types, their spectral types were assigned using spectral templates from \citet{2010ApJS..190..100K,1991ApJS...77..417K}. By using the criteria from Appendix A in \citet{2009AA...504..461F}, 308/1077 stars are classified as CTTSs. A Simbad \citep{2000A&AS..143....9W} search revealed that almost all of the TTS candidates are known photometric young stellar objects. In this work, 783/1077 stars were newly classified as CTTS/WTTS based on the strength of H$\alpha$ emission lines, among which 169 are CTTSs. All TTSs are given in Table \ref{tab: tabel1}. 
\par
There are 69 TTSs that also have LAMOST MRS. Among them, 62 stars have significant Li I absorption lines, which are marked in Table \ref{tab: tabel1}, two stars have no Li I absorption lines, and five stars cannot be determined because of the low S/N. Plots of the Li I line MRS spectra of these 69 stars are available in online data \footnote{ \url{https://doi.org/10.12149/101191} } in the rest wavelength.  Therefore, if contamination exists, the contamination rate of TTSs would be no more than $\sim$10\% (7/69). 
\par

\section{Full disk and accretion} \label{sec: sec4}

\subsection{The Full Disk}

 \begin{figure*}[ht!]
	\plotone{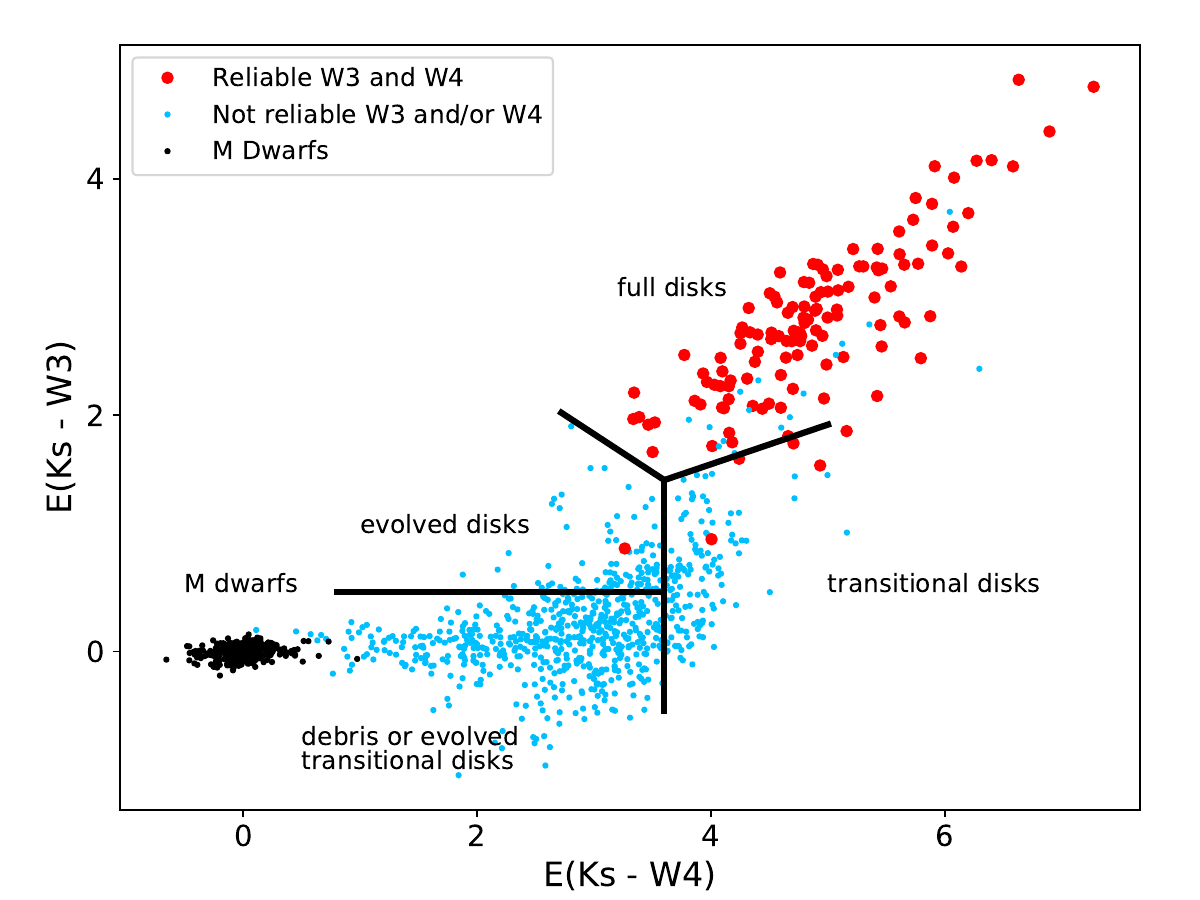}
	\caption{E(Ks-W3) vs. E(Ks-W4) for TTSs. The big red circles are TTSs with reliable W3 and W4 data, while small blue dots are TTSs without reliable W3 and/or W4 data. The black dots are M dwarfs. The boundaries of the different disk types are from \citet{2014ApJ...784..126E, 2018AJ....156...75E} and \citet{2012ApJ...758...31L}.
	\label{fig: color excess}}
\end{figure*} 

The circumstellar disk around a TTS exhibits an IR excess, which varies for different evolutionary stages. The Ks-W3 and Ks-W4 can be used to diagnose the evolutionary stage of a disk \citep{2012ApJ...758...31L}. We obtained E(Ks-W3) and E(Ks-W4) of TTSs with their intrinsic colors from \citet{2022AJ....163...24L}. In Figure \ref{fig: color excess}, the black dots are M dwarfs used in Section \ref{subsec: Photometric data},  the large red circles are TTSs that have reliable W3 and W4 data: the $ccf$ flag is ``0'' and the $qph$ flag is ``A'' for either of both bands, while the TTSs without reliable W3 and/or W4 data are marked by small blue dots. The regions of full, transitional, evolved, and debris or evolved transitional disks are also shown in Figure \ref{fig: color excess}, with the criteria from \citet{2014ApJ...784..126E, 2018AJ....156...75E} and \citet{2012ApJ...758...31L}. For red circles, except for the two outliers, all others are considered to be full disks, although a few are located around the boundary between full and transitional disks. Though the small blue dots are not reliable, it seems that these data imply that almost all TTSs are supposed to have disks. For blue dots, lacking reliable W3 and/or W4 data, the criteria of Ks-W2 excess was used to diagnose whether they have full disks as given by \citet{2012ApJ...758...31L}, \citet{2014ApJ...784..126E}, and \citet{2022AJ....163...25L}. 
\par
We checked all these full disk stars with Aladin \citep{2000A&AS..143...33B} and found that except for two stars, all other stars are not contaminated by bright infrared sources. Finally, 238 stars were confirmed as full disk stars, of which 124 were obtained by E(Ks-W3) and E(Ks-W4), and 114 were obtained by Ks-W2 excess.

\subsection{Accretion} \label{sec: accretion}

  \begin{figure*}[ht!]
  	\plotone{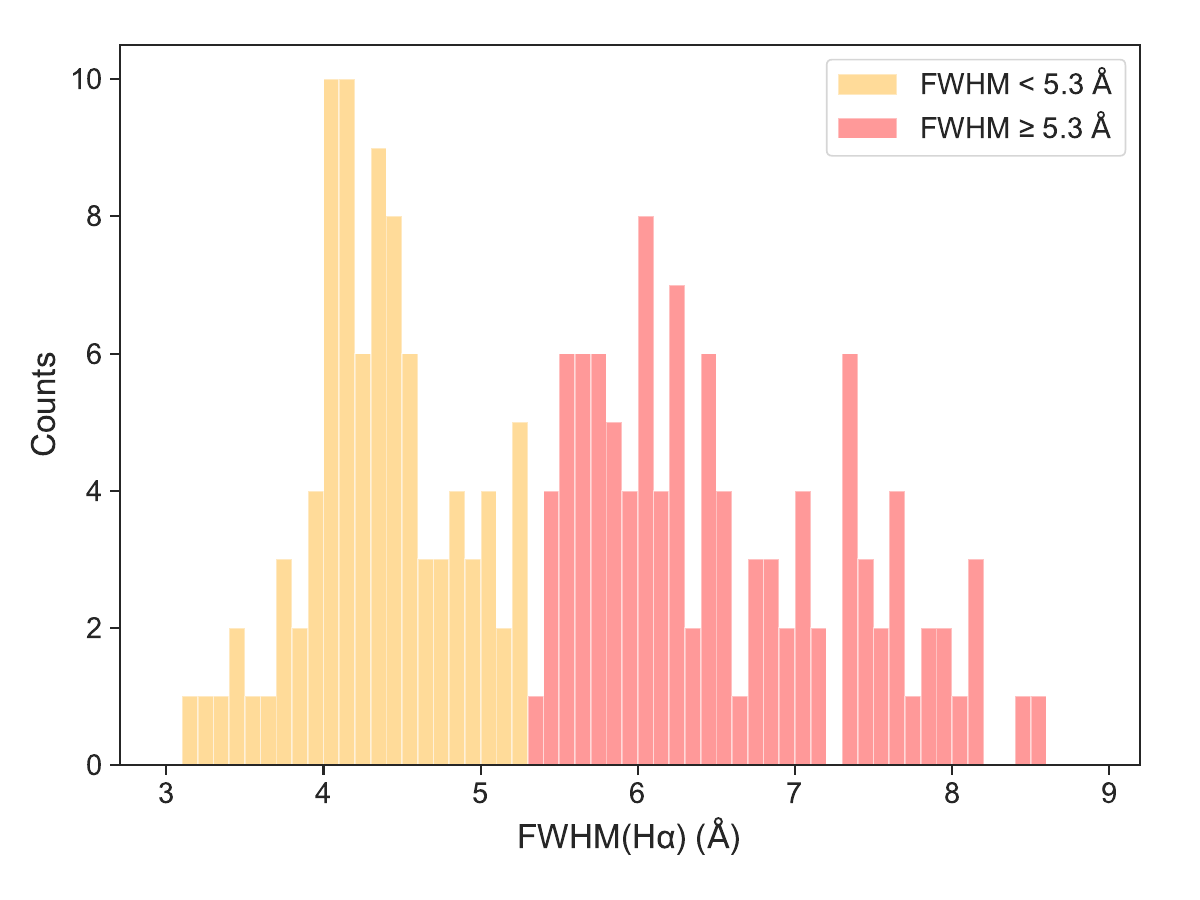}
  	\caption{Histograms of FWHMs of H$\alpha$ emission lines for TTSs with EW(H$\alpha$) $\geq$ 25 \AA.
    	\label{fig: FWHM}}
  \end{figure*}

The profile of the H$\alpha$ line can be used to obtain some information on the accretion activity. \citet{2006MNRAS.370..580K} simulated all H$\alpha$ profiles in \citet{1996A&AS..120..229R} with different combinations of the inclination, the mass-loss rate to mass-accretion rate ratio, and the wind acceleration rate. Most of the results show 10\% widths $>$ 270 km\ s$^{-1}$, a criterion defined by \citet{2003ApJ...582.1109W} indicating the full width at 10\% of the peak of the H$\alpha$ emission profile, which is independent of the spectral type, can be used to identify accretion. However, the threshold value of 10\% widths $>$ 270 km\,s$^{-1}$ is not applicable to LRS, because of the broad instrumental profile and only several flux points available to determine the 10\% width. Alternatively, we propose an FWHM method to identify accreting TTSs.
\par
To obtain EW and FWHM of a H$\alpha$ emission line, we fitted its profile in the wavelength range of $6564.61\pm20$ \AA\ with the function
  \begin{equation}
    F\left ( \lambda \right ) = A\times e^{\frac{-(\lambda - \mu )^2}{2\times\sigma^2 } }+ B\times \lambda^2 + C\times \lambda + D,
    \end{equation}
where $\lambda$ is the wavelength in angstroms, and $A$, $B$, $C$, $D$, $\mu$ and $\sigma$ are constants to be fitted. The fitted quadratic polynomial $B\times\lambda^2+C\times\lambda+D$ is used as the pseudo-continuous spectrum of the H$\alpha$. After removing the pseudo-continuous spectrum, we used the compound trapezoidal formula to calculate its EW(H$\alpha$) and performed a visual check to ensure the accuracy of the EW(H$\alpha$) and the fitted parameters. Then, the FWHM of H$\alpha$ is
  \begin{equation}
FWHM = 2 \times \sqrt{2 \times ln2} \times \sigma \approx 2.355 \times \sigma. 
  \end{equation}
\par
Figure \ref{fig: FWHM} shows a histogram of the frequency distribution of the FWHM of H$\alpha$ with the EW(H$\alpha$) $\geq$ 25 \AA, where a clear bimodal distribution separated by 5.3 \AA \ can be seen. The threshold of 25 \AA \ is chosen to ensure high S/Ns of H$\alpha$ emission lines and reduce the influence of non-full disk stars (see Figure \ref{fig: FWHM}). The FWHM of the theoretical instrumental profile of the LAMOST spectrum at 6564 \AA \ is about 3.65 \AA \ for the resolution of R = 1800. The orange data have an average FWHM of about 4.3 \AA, slightly larger than the theoretical value, and are considered to be intrinsic for the H$\alpha$ emission lines in the observed spectra, probably mainly from non-accretion activities, such as chromospheric activity and nebula emissions. Compared to this, the larger FWHM indicated by the red data may mainly originate from accretion activities. Therefore, the criterion of FWHM $\geq$ 5.3 \AA \ is used to select accreting TTSs. Furthermore, to obtain reliable accreting stars, the S/N of H$\alpha$ should be high, so we require EW(H$\alpha$) $\geq$ 15 \AA, which may result in missing some TTSs with low accretion rates. Besides, the criterion of EW(H$\alpha$) $\geq$ 15 \AA \ also means that all accreting stars here should be CTTSs.
\par
\begin{figure*}[t]
	\plotone{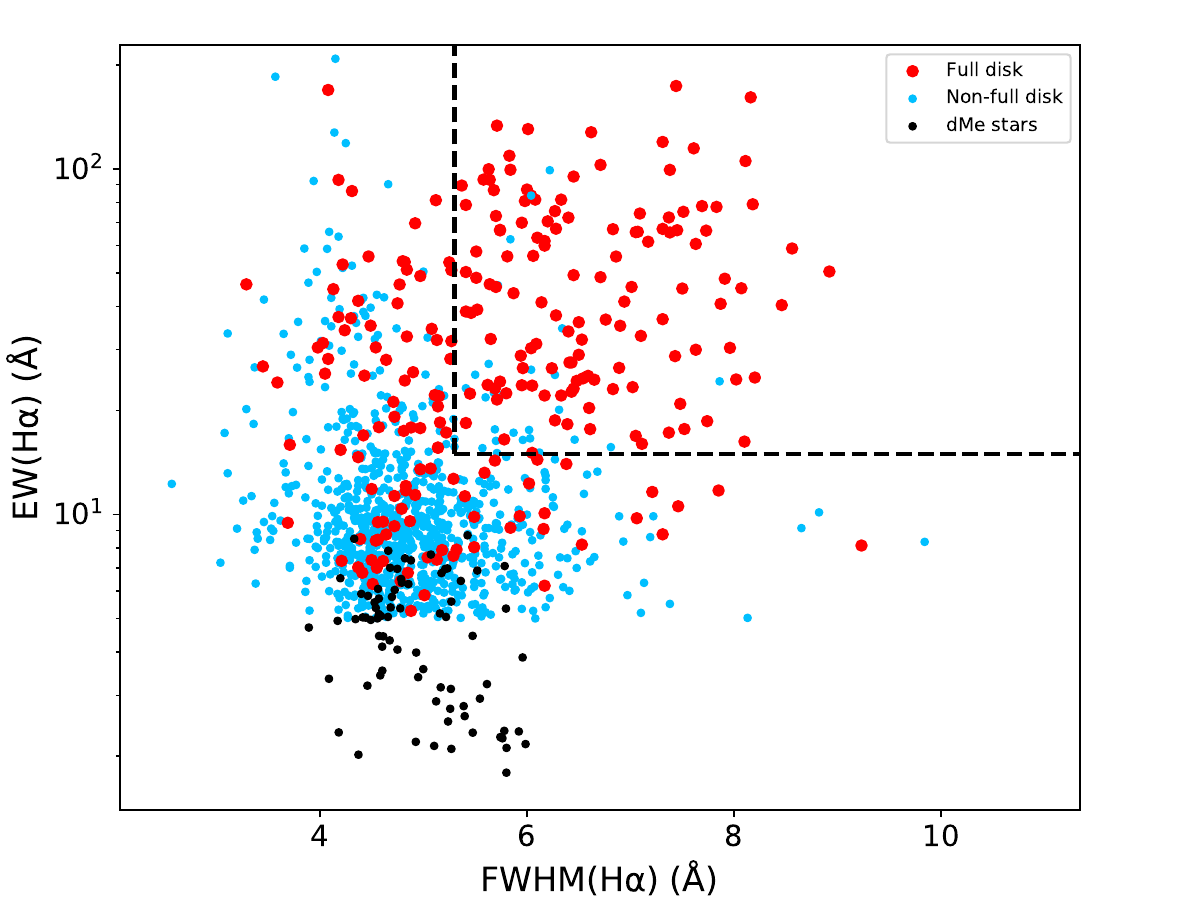}
  	\caption{EW(H$\alpha$) vs. FWHM(H$\alpha$). The red and blue circles are full-disk TTSs and non-full-disk TTSs, respectively, and the black dots are dMe stars. The accreting stars are those in the region enclosed by dotted lines: FWHM(H$\alpha$) $\geq$ 5.3 \AA \ and EW(H$\alpha$) $\geq$ 15 \AA.}
    	\label{fig: EW_FWHM}
\end{figure*}

The EW(H$\alpha$) vs. FWHM(H$\alpha$) diagram is shown in Figure \ref{fig: EW_FWHM}, and the stars in the region of FWHM $\geq$ 5.3 \AA \ and EW(H$\alpha$) $\geq$ 15 \AA \ are considered as accreting TTSs, and there are 157 in total. In Figure \ref{fig: EW_FWHM}, the dMe stars in the M dwarf samples in Section \ref{subsec: Photometric data} are shown in black dots for comparison, and they are far from accreting TTSs. Therefore, it is impossible for the accreting TTSs in Figure \ref{fig: EW_FWHM} to be contaminated by M dwarfs. 
\par
\
Only 15 accreting TTSs also have MRS in the LAMOST DR8. Their H$\alpha$ profiles show clearly asymmetry and variation because of accretion activities, which are given in online data. 
\par

\par
In Figure \ref{fig: EW_FWHM}, among the accreting TTSs, 129/157 (82\%) are full disk stars, which implies that accreting stars usually have full disks. \citet{2008AJ....136..521D} also found that 11/12 of their full disk stars are accreting.

\section{stellar parameters} \label{sec: sec5}
Stellar atmospheric parameters are needed to better understand the evolution of these TTSs. Assuming their metal abundances to be solar metal abundances, we calculated their effective temperature ($T_{\rm eff}$) and surface gravity (log $g$) using the PHOENIX templates \footnote{ \url{ http://phoenix.astro.physik.uni-goettingen.de/ } } \citep{2013A&A...553A...6H}. The PHOENIX templates are a library of high-resolution atmospheric model synthetic spectra covering the wavelength range from 50 nm to 5.5$\mu$m. The templates we used are those with [Fe/H] = 0, [$\alpha$/Fe] = 0, 2800 K $\le$ $T_{\rm eff}$ $\le$ 4200 K and 2.0 $\le$ log $g$ $\le$ 6.0. 
\par
We firstly reduce the resolution of templates from 500,000 to 1800 by convoluting a Gaussian function and then obtain new templates with a $\triangle T_{\rm eff} = 20 $ K and $\triangle \log g = 0.1$ dex by linear interpolations, which were used to fit LAMOST spectra to calculate $T_{\rm eff}$ and $\log g$.
\par
We used the following model to fit the LAMOST LRS of TTSs:
  \begin{equation}
    \label{equ:obspec}
    F_O( \lambda) =P_{m} ( \lambda) \times F_T( \lambda) + C
  \end{equation}
where $\lambda$ is the wavelength in \AA, $F_O( \lambda)$ is a LAMOST spectrum, $F_T( \lambda)$ is a template, $P_m(\lambda)$ is a polynomial with an order of $m$, and $C$ is a constant to represent the veiling caused by accretion. We found that $m = 1$ is good enough for most stars, so a linear function $P_1(\lambda)$ is used for most stars, while higher orders are used only for a few stars. The presence of an accretion continuum will reduce the depth of the photospheric absorption lines and a CTTS spectrum can be represented as a photospheric spectrum with an addition of a veiling spectrum \citep{1989ApJS...70..899H, 2003ApJ...583..334H}. \citet{2014ApJ...786...97H} suggested that the optical accretion continuum is flat and can be approximated by a constant. Therefore, the constant $C$ is adopted in this model. At last, all spectra with their models are checked by eye. The precisions of the parameters are $\sigma_{T_{\rm eff}}$ = 23.96 K and $\sigma_{\log g}$ = 0.12 dex, using the same method as \citet{2021RAA....21..202D}, 
as shown in the online data.

\begin{deluxetable*}{rrrrccrcrrcccccc}
\tablecaption{M-type TTS catalog \label{tab: tabel1}}
		\tablewidth{700pt}
		\tabletypesize{\scriptsize}
		\tablehead{
			\colhead{Seq} &\colhead{GLON} &\colhead{GLAT} & \colhead{$T_{\rm eff}$} & \colhead{logg} & \colhead{logAge} & \colhead{Mstar} & \colhead{SpT} & \colhead{EWHa} & \colhead{FWHM} & \colhead{Li I$^{a}$} & \colhead{TTS$^{b}$} & \colhead{Acc$^{c}$} & \colhead{Disk$^{d}$} & \colhead{New$^{e}$} & \colhead{...}\\
			\colhead{} & \colhead{(deg)} & \colhead{(deg)} & \colhead{(K)} & \colhead{(cm\,s$^{-2}$)} & \colhead{(yr)} &\colhead{$M_{\sun}$} & \colhead{} & \colhead{(\AA)} & \colhead{(\AA)} & \colhead{ } &\colhead{} & \colhead{ } & \colhead{ } & \colhead{ }& \colhead{} 
		}						
		\startdata
            1	&	178.162823	&	-6.218360	&	3004	&	3.69	&	6.41 	&	0.21 	&	M5	&	14.67	&	5.80	&	-	&	W	&	Y	&	-	&	Y & ...\\
            2	&	171.361519	&	-10.774522	&	3465	&	2.28	&	3.95 	&	0.25 	&	M5	&	6.34	&	4.60	&	Y	&	W	&	-	&	-	&	Y & ...\\
            3	&	173.583527	&	-12.373890	&	3095	&	4.26	&	7.24 	&	0.32 	&	M5	&	11.62	&	5.24	&	-	&	W	&	-	&	-	&	Y & ...\\
            4	&	173.676373	&	-8.037127	&	3274	&	3.12	&	5.52 	&	0.22 	&	M6	&	9.73	&	5.91	&	-	&	W	&	Y	&	-	&	Y & ...\\
            5	&	171.120195	&	-9.801903	&	3001	&	3.79	&	6.54 	&	0.24 	&	M5	&	9.79	&	4.40	&	-	&	W	&	-	&	-	&	Y	 & ...\\
            6	&	171.480895	&	-10.625092	&	3379	&	2.36	&	3.95 	&	0.22 	&	M5	&	9.50	&	4.56	&	Y	&	W	&	-	&	Y	&	-	&	 ...
		\enddata
		\tablecomments{\\
			a ``Y'' means obvious Li I absorption line; ``N'' means no Li I absorption line; ``?'' means it cannot be determined because of the low S/N; ``-'' means that there is no MRS. \\
			b The TTS type is given according to \citet{2009AA...504..461F}: ``C'' for CTTS and ``W'' for WTTS.\\
			c ``Y'' means star in accretion.\\
			d ``Y'' for a full disk.\\
			e ``Y'' for the TTS newly identified by spectra.
        \tablecomments{Table \ref{tab: tabel1} is published in its entirety in the machine-readable format
            in the online Journal and on China-VO: \dataset[https://doi.org/10.12149/101191]{https://doi.org/10.12149/101191}. 
            A portion is shown here for guidance regarding its form and content.}
		}
	\end{deluxetable*}

\section{Evolution of full disks} \label{sec: sec6}
\subsection{The Stellar Evolution Model}
For the analysis of the disk evolution of TTSs, the PARSEC evolutionary model \footnote{PARSEC version 1.2S: \url {http://stev.oapd.inaf.it/cgi-bin/cmd} } \citep{2012MNRAS.427..127B, 2014MNRAS.444.2525C,2014MNRAS.445.4287T} was used, because the PARSEC model uses the $T-\tau$ relations from the PHOENIX BT-Settl model atmosphere as the outer boundary conditions, so the PARSEC evolutionary model can give self-consistent stellar evolutionary stages from the stellar atmosphere parameters calculated in this paper.
\par
The stellar ages and masses were calculated from the PARSEC evolutionary grids by linear interpolations, as shown in Figure \ref{fig: H-R  diagram}. The maximum errors of $T_{\rm eff}$ and log $g$ are the grid intervals of the PHOENIX templates, which are 100 K for $T_{\rm eff}$ and 0.5 dex for log $g$, respectively, as shown in the left corner of 
Figure \ref{fig: H-R diagram}.
\par
Figure \ref{fig: H-R diagram} shows the Hertzsprung-Russell (H-R) diagram of TTSs. The blue empty circles are accreting TTSs, and 120/157 (76\%) accreting TTSs are younger than 1 Myr, implying that most accretion activities happen in the first 1 Myr, consistent with the finding of \citet{2004AJ....128..805S}. Several accreting TTSs are as old as about 20 Myr, which may be Peter Pan disk candidates.

\subsection{Two Different Evolutionary Patterns of Full Disks } \label{sec: Two Different Evolutionary Patterns of Full Disks}
\begin{figure*}[t]
	\plotone{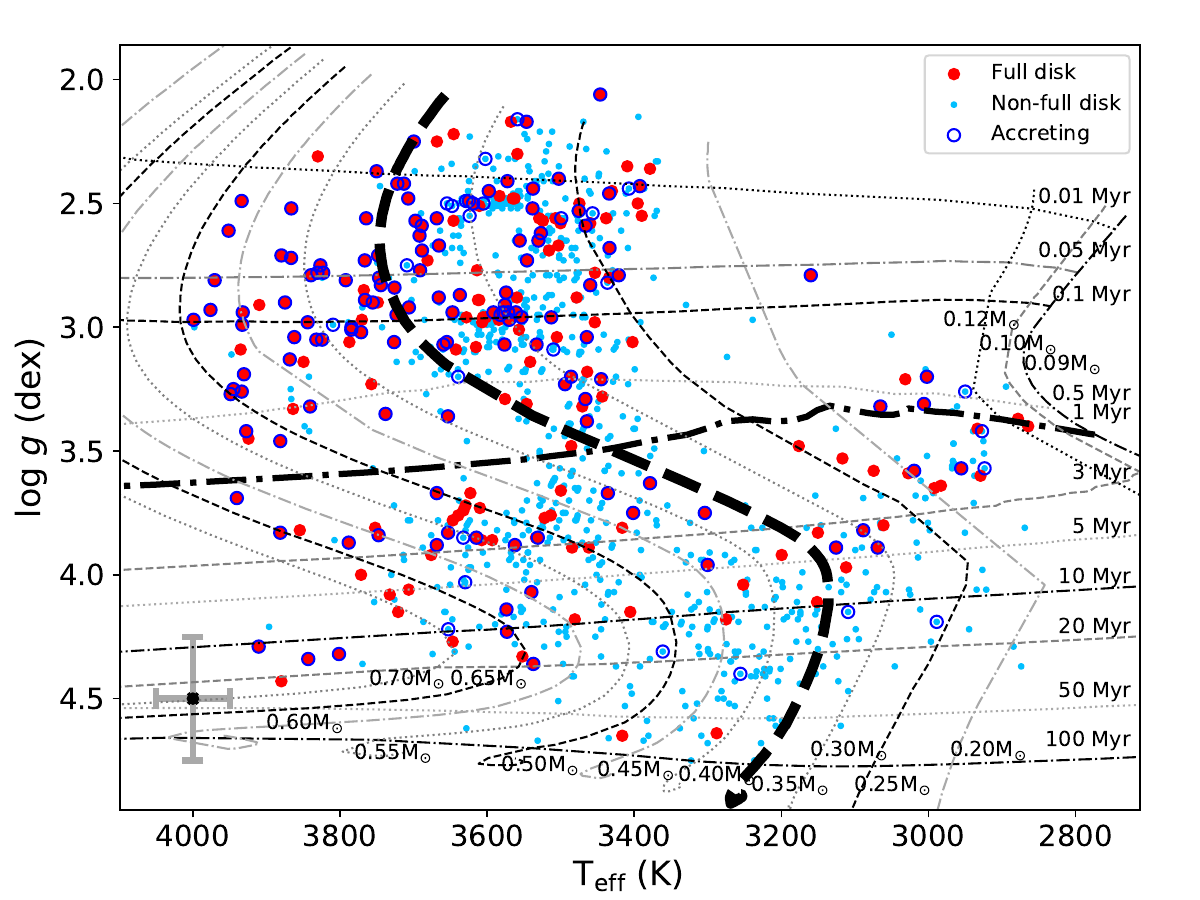}
  	\caption{The H-R diagram of TTSs with age and mass isochrones. Red full circles are stars with full disks, blue full circles are stars with non-full disks, 
	and empty circles are stars with accretions. The maximum errors of $T_{\rm eff}$ and $\log g$ are shown in the left corner.}
    	\label{fig: H-R diagram}
\end{figure*}

\begin{figure*}[ht]
          \centering
    \subfigure{			
\includegraphics[width=0.45\textwidth,trim=0 0 0 0,clip]{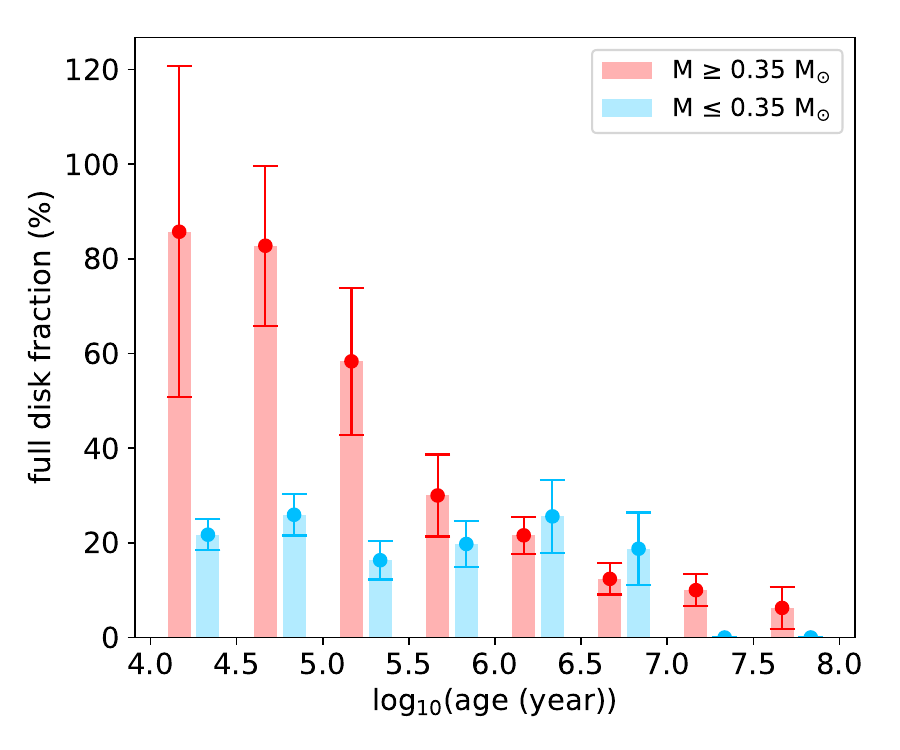}}
    \subfigure{			\includegraphics[width=0.45\textwidth,trim=0 0 0 0,clip]{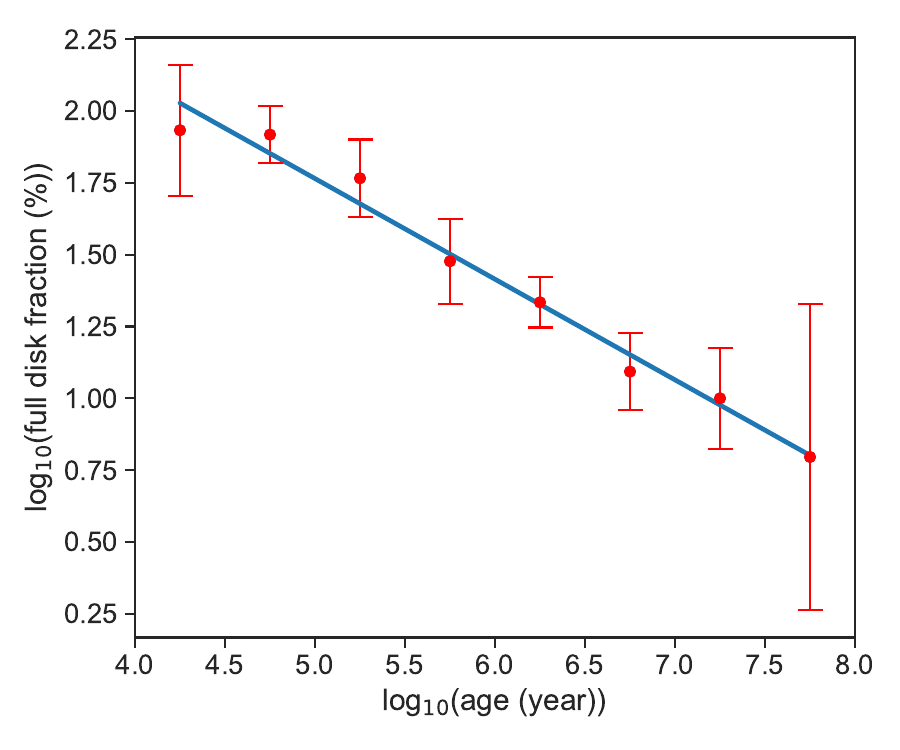}}
    \caption{The distributions of the fraction of full-disk TTSs. The right panel is for $M > 0.35 \, M_{\sun}$ and the blue line is Equation \ref{equ: frac}.
    \label{fig: full-disk fraction}}
\end{figure*}

\begin{figure*}[ht]
          \centering
    \subfigure{			
\includegraphics[width=0.45\textwidth,trim=0 0 0 0,clip]{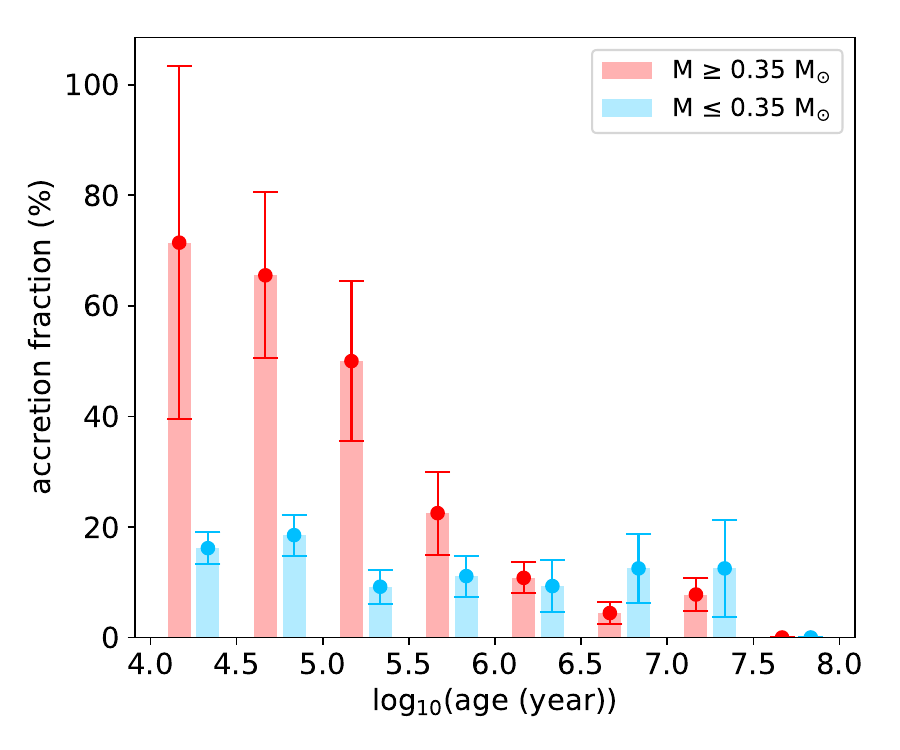}}
    \subfigure{			\includegraphics[width=0.45\textwidth,trim=0 0 0 0,clip]{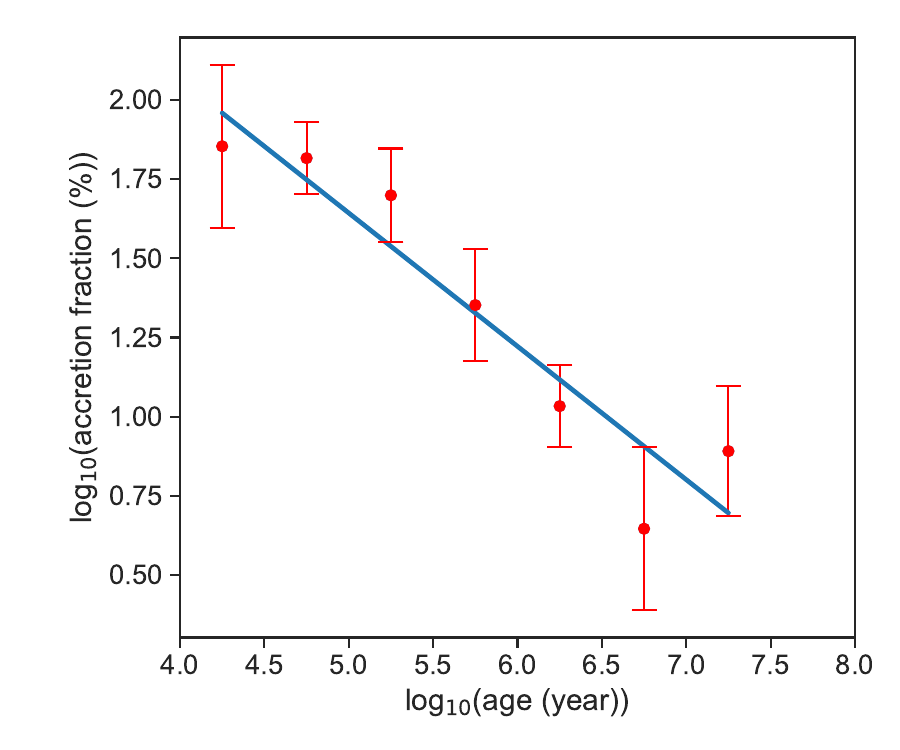}}
    \caption{The distributions of the fraction of accreting TTSs. The right panel is for $M > 0.35 \, M_{\sun}$ and the blue line is Equation \ref{equ:acc}. 
    \label{fig: accretion fraction}}
\end{figure*}

\begin{figure*}[ht]
          \centering
    \subfigure{			
\includegraphics[width=0.45\textwidth,trim=0 0 0 0,clip]{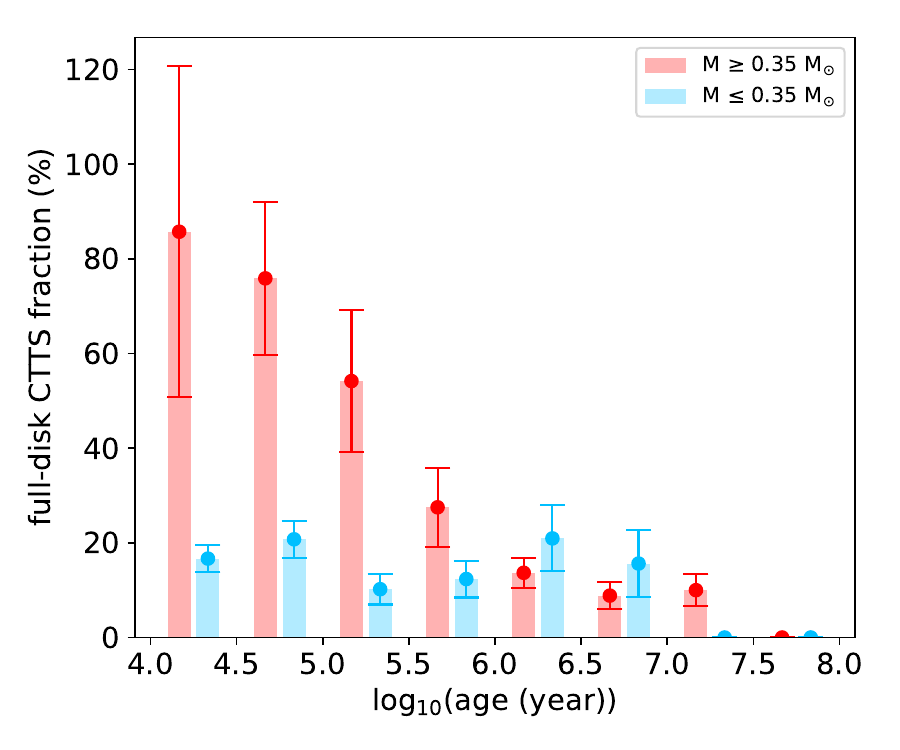}}
    \subfigure{			\includegraphics[width=0.45\textwidth,trim=0 0 0 0,clip]{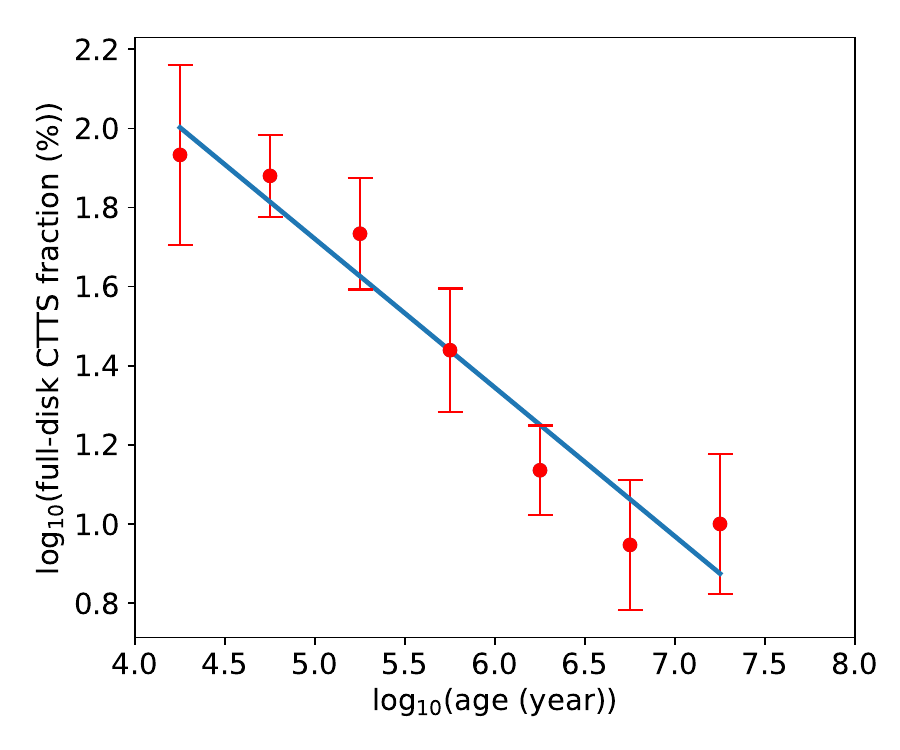}}
    \caption{The distributions of the fraction of full-disk CTTSs. The right panel is for $M > 0.35 \, M_{\sun}$, and the blue line is Equation \ref{equ:CTTS}. 
    \label{fig: CTTS fraction}}
\end{figure*}

\par

Separated by $M = 0.35 M_{\sun}$, the full-disk TTSs show two different evolutionary patterns, which are shown by the full-disk fraction in Figure \ref{fig: full-disk fraction}, and the accretion fraction of TTSs also shows two different evolutionary patterns, as shown in Figure \ref{fig: accretion fraction}. Because almost all our TTSs have disks, the full-disk fraction here is the ratio of full-disk TTSs to disk-bearing TTSs, and the accretion fraction is the ratio of accreting TTSs to disk-bearing TTSs.
\par
In Figure \ref{fig: full-disk fraction}, the full-disk fractions of TTSs with $M > 0.35 \, M_{\sun}$ and $M < 0.35 \, M_{\sun}$ are shown in the left panel by the red and blue histogram, respectively. We can see that the full-disk fraction of $M > 0.35 \, M_{\sun}$ declines with age, while the full-disk fraction of $M < 0.35 \, M_{\sun}$ remains at a constant of $\sim$20\% until $\sim$5 Myr.  The fraction of full-disk TTSs for $M > 0.35 \, M_{\sun}$ is fitted in the logarithmic plane and shown in the right panel of Figure \ref{fig: full-disk fraction}, which is
\begin{equation}
\label{equ: frac}
\log_{10}(f_{d}) = -0.35 (\pm 0.02) \log_{10}(t) + 3.52 (\pm 0.12)
\end{equation}
where $f_{d}$ is the fraction of full disk stars for $M > 0.35 \, M_{\sun}$ in percentage and $t$ is the stellar age in years. 
\par

In Figure \ref{fig: accretion fraction}, the accretion fraction of TTSs with $M > 0.35 \, M_{\sun}$ and $M < 0.35 \, M_{\sun}$ are shown in the left panel by the red and blue histogram, respectively. For TTSs with $M > 0.35 \, M_{\sun}$, the accretion fraction declines with age, while the accretion fraction of $M < 0.35 \, M_{\sun}$ remains a constant of $\sim$10\% until $\sim$5 Myr with two outliers longer than 10 Myr (see Figures \ref{fig: H-R diagram} and \ref{fig: accretion fraction}).  The accretion fraction of TTSs with $M > 0.35 \, M_{\sun}$ can be fitted as
\begin{equation}
\label{equ:acc}
\log_{10}(f_{\rm acc}) = -0.42 (\pm 0.07) \log_{10}(t) + 3.75 (\pm 0.39)    
\end{equation}
where $f_{\rm acc}$ is the accretion fraction for $M > 0.35 \, M_{\sun}$ in percentage, and $t$ is the stellar age in years. 

\par
From Equations \ref{equ: frac} and \ref{equ:acc} we can see that for TTSs with $M > 0.35 \, M_{\sun}$, 
both the full-disk fraction and the accretion fraction show similar declining trends with age, with the accretion fraction appearing to decline slightly faster, even though the rates overlap within the fit uncertainties.
Equation \ref{equ: frac} shows that the full disks evolve faster than the non-full disks, and after 1 Myr, the disks are dominated by non-full-disk TTSs (Figure \ref{fig: H-R diagram}). Equation \ref{equ:acc} of the accretion fraction shows a similar tendency to that of Equation \ref{equ: frac}, which suggests that most accreting TTSs also have full disks (see Figure \ref{fig: EW_FWHM}), and accretion activities disappear with the disappearance of full disks. 
\par
The LRS of LAMOST can only find strong accreting stars by using the criteria in Section \ref{sec: accretion} but miss weak accreting TTSs. Now, we assume that CTTSs with full disks are all accreting stars. The full-disk CTTS fraction with $M > 0.35 \, M_{\sun}$ and $M < 0.35 \, M_{\sun}$ are shown in the left panel by the red and blue histograms, respectively, in Figure \ref{fig: CTTS fraction}, from which we can see that there are also two different evolutionary patterns: the full-disk CTTS fraction with $M > 0.35 \, M_{\sun}$ declines with age, while remains a constant of $\sim$ 15\% for $M < 0.35 \, M_{\sun}$. 
\par
The full-disk CTTS fraction with $M > 0.35 \, M_{\sun}$ can be fitted as
\begin{equation}
\label{equ:CTTS}
\log_{10}(f_{\rm CTTS}) = -0.38 (\pm 0.04) \log_{10}(t) + 3.60 (\pm 0.25)    
\end{equation}
where $f_{\rm CTTS}$ is the full-disk CTTS fraction for $M > 0.35 \, M_{\sun}$ in percentage, and $t$ is the stellar age in years.
\par
\begin{figure*}[ht!]
      \centering
    \subfigure{			
    \includegraphics[width=0.95\textwidth,trim=0 0 0 0,clip]{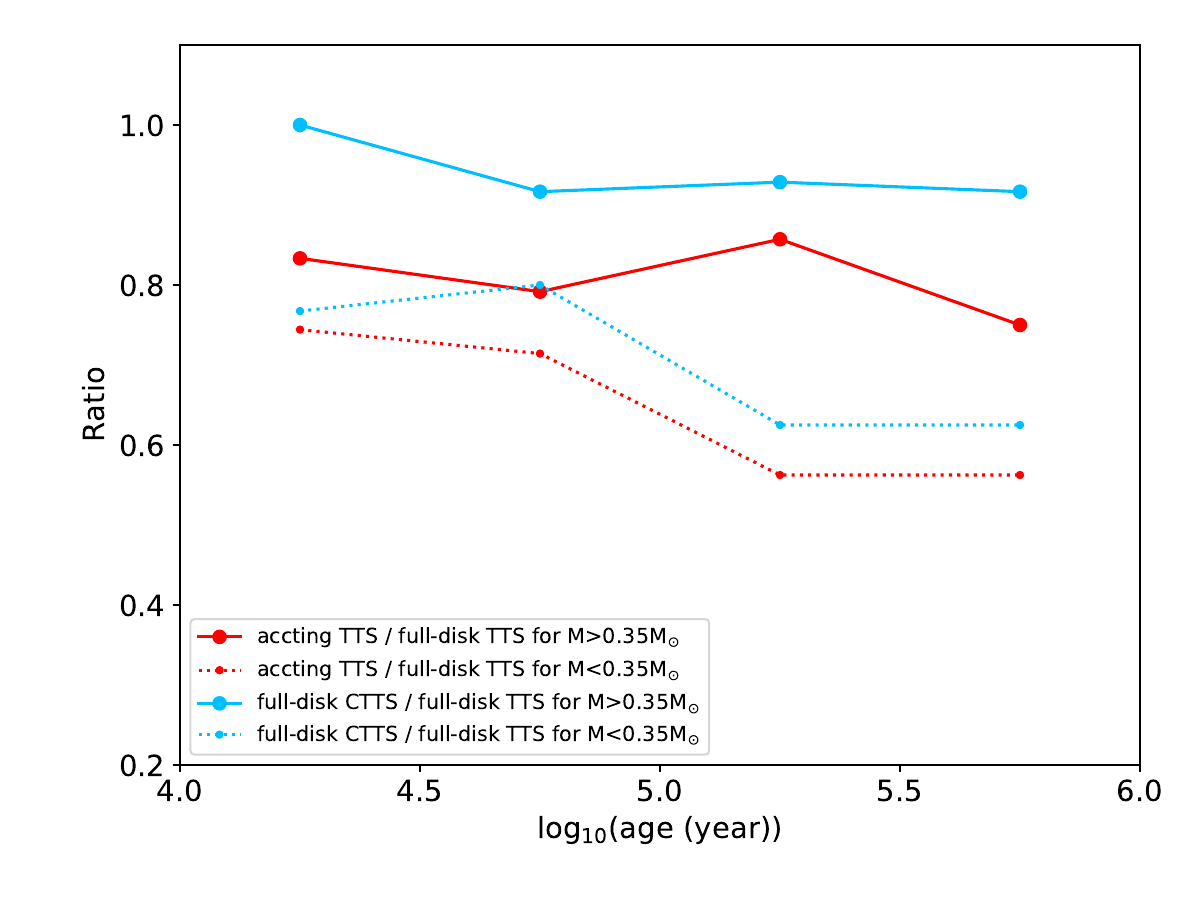}}
    \caption{The ratios of accreting TTSs or full-disk CTTSs to the full-disk TTSs within 1 Myr. The red lines indicate the ratios of the accreting TTSs to the full-disk TTSs, while the blue lines indicate the ratios of the full-disk CTTSs to the full-disk TTSs. The solid lines are for $M > 0.35 \, M_{\sun}$ and the dotted lines are for $M < 0.35 \, M_{\sun}$.
          \label{fig: ratio}}
  \end{figure*} 

When comparing the ratio of accreting TTSs to full-disk TTSs as a function of age between two mass groups within 1Myr as shown in Figure \ref{fig: ratio}, this ratio appears to be systematically lower for $M < 0.35 \, M_{\sun}$ than for $M > 0.35 \, M_{\sun}$. It also happens in comparing the ratio of the full-disk CTTSs to the full-disk TTS. This interesting effect may be explained by the dependence of the accretion rate on the stellar mass \citep{2022AJ....163...74T}, with lower-mass stars tending to have lower accretion rates and being more difficult to detect by the H$\alpha$ line in the low-resolution data. The mass dependence of accretion fractions of TTSs shown in Figure \ref{fig: ratio} may be reflective of an observational bias in the detection of accretion levels, with the lower-mass stars crossing below the detection threshold earlier than higher-mass stars.
\par
We also checked the MIST model \footnote{ \url{ http://waps.cfa.harvard.edu/MIST/model_grids.html\#isochrones} } \citep{2016ApJ...823..102C}, which has near-vertical isomass lines compared to the S-shaped isomass lines of the PARSEC model. The MIST model does not deny the existence of two full disk evolutionary patterns either. For TTSs with $M > 0.4 \, M_{\sun}$ and ages $<$ 1 Myr, 57/80 (71\%) stars have full disks, while the percentage drops to 111/534 (21\%) for TTSs with $M < 0.4 \, M_{\sun}$ in the same age range. For TTSs with $M > 0.4 \, M_{\sun}$, the fraction of full-disk TTSs to disk-bearing TTSs seems to decline with age following a relationship of $f \propto t^{-0.35\pm0.08}$, implying that their full disks evolve faster than non-full disks, while for TTSs with $M < 0.4 \, M_{\sun}$, they have a stable full-disk fraction of $\sim$16\% and seem to evolve at similar rates as non-full disks. The H-R diagram and the distribution of full-disk fraction with age under the MIST model are available in the online data.

\subsection{Two evolutionary patterns in individual star-forming regions} \label{subsec: 6.3}

\begin{figure*}[ht!]
      \centering
    \subfigure{			\includegraphics[height=0.95\textwidth, width=0.95\textwidth,trim=0 0 0 0,clip]{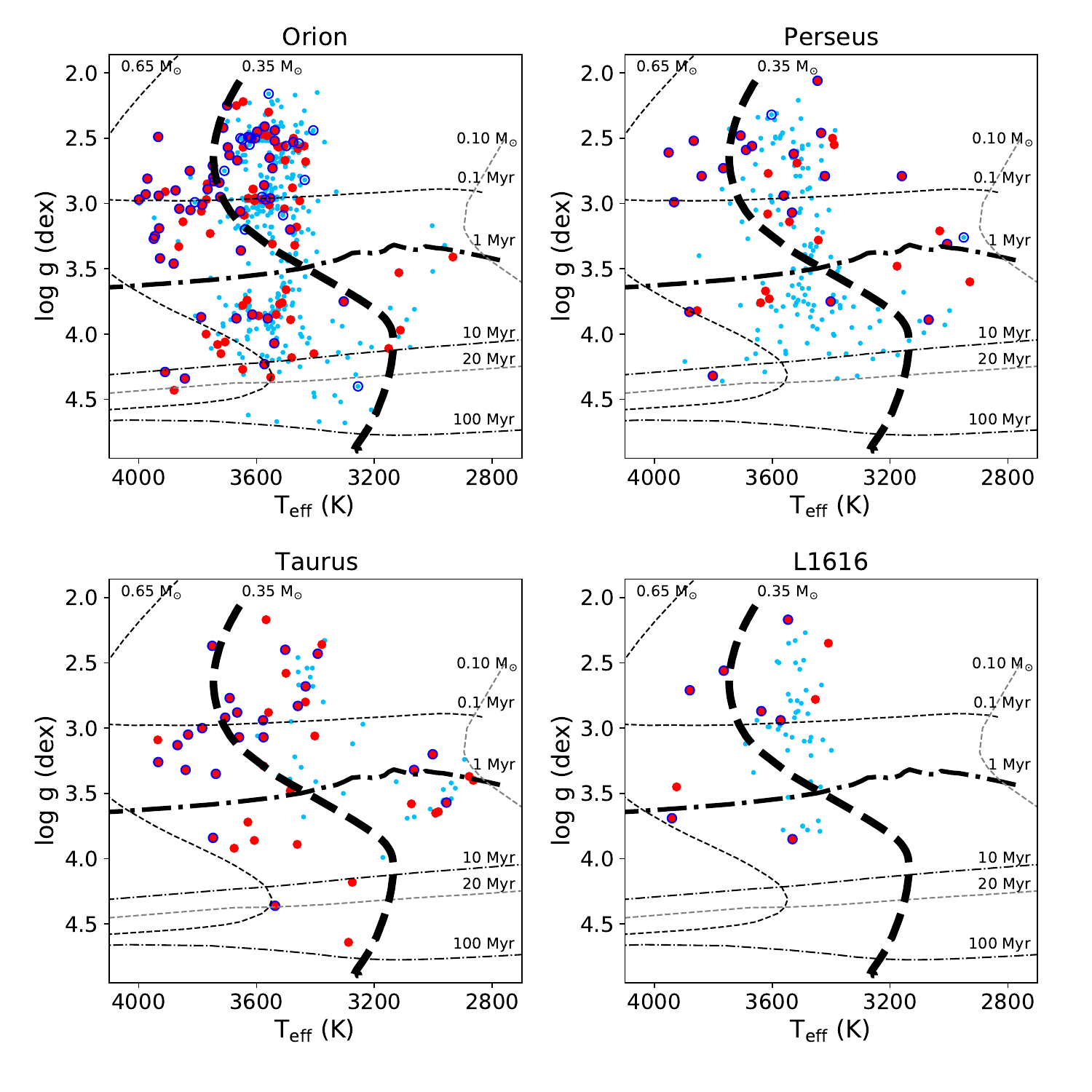}}
    \caption{Same as Figure \ref{fig: H-R diagram}, but for individual star-forming regions.
          \label{fig: star-forming region}}
  \end{figure*} 
 
\par
To check if the real existence of the above two different evolutionary patterns is just specific to a certain star-forming region, we inspected the H-R diagrams of well-studied star-forming regions with more than 50 candidate member stars, which are Orion, Perseus, Taurus, and L1616, as shown in Figure \ref{fig: star-forming region}. We found that it seems that none of these star-forming regions can reject our suggestion of the two different evolutionary patterns. Furthermore, the ages of these 4 star-forming regions are consistent with those given in the literature: $<$ 10 Myr for Orion \citep{2017AJ....153..188F, 2022MNRAS.517..161K}, $<$ 15 Myr for Perseus \citep{2015AJ....150...95A}, 1-3 Myr for Taurus \citep{2023AJ....165...37L}, and 1-2 Myr for L1616 \citep{2004A&A...416..677A}. In other words, our results seem reliable. 

\section{Discussion} \label{sec: sec7}
The timescale of the full disk is very important for the formation of planets, especially for giant planets. In Figure \ref{fig: H-R diagram}, almost all stars with $M>0.35 M_{\odot}$ are born with full disks, and most of them can survive for a few 0.1 Myr, which is about the timescale of planetesimal formation revealed by meteorites: several 0.1 Myr \citep{2022ASSL..466....3R}. Thus, the planetesimals may have been formed in the full disks. We notice that some full disks can even survive for as long as 20 Myr. If this is true, some TTSs with $M > 0.35 \, M_{\sun}$ have a long time to form planets, which may result in the diversity of planet development. 
\par

For TTSs with $M<0.35 M_{\odot}$, from Figures \ref{fig: full-disk fraction} and  \ref{fig: H-R diagram}, 
we can see that as many as $\sim$80\% TTSs seem to already be deprived of full disks at ages $\sim$0.1 Myr and almost all of them have no obvious accretion activities, which implies that they are unable to grow up by accretion, thus of low masses when they become dwarfs, while only $\sim$20\% of TTSs still have full disks at 0.1Myr, and this ratio can remain for $\sim$5 Myr, which suggests that full disks evolve at a similar rate to non-full disks in the first 5 Myr, then disappear and only non-full disks are left. These long-lived full disks can provide copious gas and dust for the formation of planets, especially giant planets, so the giant planets might be formed therein (see Figure 8 in \citet{2020plas.book..325M}). GJ 3512 has a mass of 0.123$\, M_{\odot}$, with a planet of $M > 0.463 \, M_{\rm Jup}$ \citep{2019Sci...365.1441M}, which might be formed in a massive long-lived full disk. 
\par
In this work, we found that for TTSs with $M>0.35 M_{\odot}$, full disks evolve faster than non-full disks (Equation \ref{equ: frac}), and accretion activities disappear with the disappearance of full disks, which suggests that with disappearance of full disk, the inner disk also disappears and thus the accretion activities stop, but the outer disk can still exist for a longer time. This supports the inside-out disk-clearing scenarios as found in \citet{2014A&A...561A..54R}. However, for TTSs with $M < 0.35 M_{\odot}$, full disks and accretions disappear at a similar rate as non-full disks in the first 5 Myr, and at 10 Myr all full disks have disappeared while non-full disks still exist, which suggests that an inside-out disk-clearing mechanism may still be operating, but on longer timescales.
\par
The difference between these two cases on the inside-out disk clearing timescale
might result from the photoevaporation: the more massive the central star the stronger the disk wind and then the faster the inner disk disappears \citep{2021ApJ...910...51K,2021MNRAS.508.3611P}. For TTSs with $M > 0.35 M_{\odot}$, their disk winds are strong enough to blow away inner disks, but the disk winds of TTSs with $M < 0.35 M_{\odot}$ are too weak to blow away inner disks and their inner disks can evolve with a similar rate as outer disks for a long time.
\par
The $0.35 \, M_{\sun}$ is a very interesting threshold, at which stellar models predict a transition from partial to fully convective interior structure for stars \citep[e.g.][]{1997A&A...327.1039C}. In fact, \citet{2018ApJ...861L..11J} found a gap around this position and suggested that it should be related to the onset of full convection. In this paper, we also found the two different evolutionary patterns of M-type TTS disks separated by this threshold. Therefore, the two different evolutionary patterns of full disks might be consequential for models of both early stellar evolution and planet formation.

\section{summary} \label{sec: sec8}
To study the circumstellar disks around M-type TTSs where the planets form, we searched and identified 1077 M-type TTSs in star-forming regions with spectra of the LAMOST DR8, and obtained the following findings:
\begin{itemize}
\item With the help of the photometric data from 2MASS, WISE, and Gaia, and the parallaxes provided by Gaia, we constructed a 36-dimensional feature vector. Then, we used the triplet network algorithm to search for TTS candidates in the LAMOST DR8. 
\item We required stars should be located in star-forming regions and their EW($\alpha$) $\geq$ 5 \AA. Finally, 1077 reliable TTS candidates were identified, of which 308 are CTTSs. After crossmatching with Simbad, a total of 783 TTSs are newly identified by spectra, including 169 new CTTSs. 
\item With the help of 2MASS and WISE data, we found that almost all 1077 TTSs have disks, of which 238 TTSs have full disks. 
\item Using the criteria of EW(H$\alpha$)$\geq$ 15 \AA \  and FWHM (H$\alpha$) $\geq$ 5.3 \AA, 157 accreting TTSs were found.
\item $T_{\rm eff}$, $\log g$ of all 1077 TTSs were calculated from LAMOST LRS, from which their ages and masses were derived.
\item About 82\% (129/157) accreting TTSs also have full disks. 
\item Most accretion activities (120/157 $\approx$ 76\%) happen in the first 1 Myr.
\item For TTSs with $M > 0.35 \, M_{\sun}$, almost all of their full disks can survive more than 0.1 Myr, and some can even last 20 Myr. The fraction of full-disk TTSs seems to decline with age following a relationship of $f \propto t^{-0.35}$, which implies that full disks evolve faster than non-full disks.
\item For TTSs with $M<0.35 \, M_{\sun}$, $\sim$80\% stars seem to lose their full disks already at ages $\sim$0.1 Myr, which may explain their lower mass.

\item For TTSs with $M<0.35M_{\sun}$, the remaining $\sim$20\% have full disks, and this fraction can persist for $\sim$5 Myr and then disappear suddenly, which implies the full disks evolve at a similar rate as non-full disks. Because of the long life of full disks of TTSs with $M<0.35 \, M_{\sun}$, the giant planets might be formed therein.
\item For TTSs of either $M < 0.35 \, M_{\sun}$ or $M > 0.35 \, M_{\sun}$, as seen from the accretion fraction and the full-disk CTTS fraction, the evolution of accretion activities is similar to that of the full disks within uncertainties, but it seems that accretion evolves slightly faster than the full disks.
\item For TTSs with $M < 0.35 \, M_{\sun}$, the inside-out disk clearing mechanism may still be operating, but on longer timescales.

\item For full disk stars, the ratio of accretion of lower-mass stars is systematically lower than that of higher-mass stars, confirming the dependence of accretion on stellar mass, which may be reflective of an observational bias in the detection of accretion levels, with the lower-mass stars crossing below the detection threshold earlier than higher-mass stars.
\end{itemize}
\par
We provide some additional material in the form of online data, all downloadable at the DOI \url{https://doi.org/10.12149/101191}, including the catalog of 1077 TTSs in txt format (the full version of Table \ref{tab: tabel1}), as well as some figures in pdf format: the MRS spectra of the 69 Li I lines mentioned in Section \ref{sec: sec3}, 
the MRS spectra of the H$\alpha$ lines of the 15 accreting TTSs discussed in Section \ref{sec: sec4}, 
the precision of the parameters mentioned in Section \ref{sec: sec5}, 
the H-R diagram of CTTSs, the H-R diagram under the MIST model, and the distribution of full-disk fractions with age under the MIST model bounded by 0.4 M$_{\sun}$ all mentioned in Section \ref{sec: sec6}, and the stellar parameter fitting results of all 1077 TTSs.

\begin{acknowledgments}		
\par
The authors thank the anonymous referees for very helpful comments, and Dr. Min Fang for personal discussions.
\par
We acknowledge the science research grants from the China Manned Space Project with No. CMS-CSST-2021-A08, and the National Natural Science Foundation of China (NSFC) with No. 12073038.
\par
Guoshoujing Telescope (the Large Sky Area Multi-Object Fiber Spectroscopic Telescope LAMOST) is a National Major Scientific Project built by the Chinese Academy of Sciences. Funding for the project has been provided by the National Development and Reform Commission. LAMOST is operated and managed by the National Astronomical Observatories, Chinese Academy of Sciences.
\par
Data resources are supported by the China National Astronomical Data Center (NADC) and the Chinese Virtual Observatory (China-VO). This work is supported by Astronomical Big Data Joint Research Center, co-founded by the National Astronomical Observatories, Chinese Academy of Sciences and Alibaba Cloud.
\par
This work has made use of data from the European Space Agency (ESA) mission {\it Gaia} (\url{https://www.cosmos.esa.int/gaia}), processed by the {\it Gaia} Data Processing and Analysis Consortium (DPAC \url{https://www.cosmos.esa.int/web/gaia/dpac/consortium}). Funding for the DPAC has been provided by national institutions, in particular, the institutions participating in the {\it Gaia} Multilateral Agreement.
\par
This research has made use of the SIMBAD database, operated at CDS, Strasbourg, France.
\par
This publication makes use of data products from 2MASS, which is a joint project of the University of Massachusetts and the Infrared Processing and Analysis Center/California Institute of Technology, funded by the National Aeronautics and Space Administration and the National Science Foundation.
\par
This publication makes use of data products from WISE, which is a joint project of the University of California, Los Angeles, and the Jet Propulsion Laboratory/California Institute of Technology, funded by the National Aeronautics and Space Administration.	
\end{acknowledgments}
\bibliography{my_paper}{}

\begin{thebibliography}{}
\expandafter\ifx\csname natexlab\endcsname\relax\def\natexlab#1{#1}\fi
\providecommand{\url}[1]{\href{#1}{#1}}
\providecommand{\dodoi}[1]{doi:~\href{http://doi.org/#1}{\nolinkurl{#1}}}
\providecommand{\doeprint}[1]{\href{http://ascl.net/#1}{\nolinkurl{http://ascl.net/#1}}}
\providecommand{\doarXiv}[1]{\href{https://arxiv.org/abs/#1}{\nolinkurl{https://arxiv.org/abs/#1}}}

\bibitem[{{Alcal{\'a}} {et~al.}(2004){Alcal{\'a}}, {Wachter}, {Covino},
  {Sterzik}, {Durisen}, {Freyberg}, {Hoard}, \&
  {Cooksey}}]{2004A&A...416..677A}
{Alcal{\'a}}, J.~M., {Wachter}, S., {Covino}, E., {et~al.} 2004, \aap, 416,
  677, \dodoi{10.1051/0004-6361:20034495}

\bibitem[{{Alexander} {et~al.}(2014){Alexander}, {Pascucci}, {Andrews},
  {Armitage}, \& {Cieza}}]{2014prpl.conf..475A}
{Alexander}, R., {Pascucci}, I., {Andrews}, S., {Armitage}, P., \& {Cieza}, L.
  2014, in Protostars and Planets VI, ed. H.~{Beuther}, R.~S. {Klessen}, C.~P.
  {Dullemond}, \& T.~{Henning}, 475,
  \dodoi{10.2458/azu_uapress_9780816531240-ch021}

\bibitem[{{Ansdell} {et~al.}(2017){Ansdell}, {Williams}, {Manara}, {Miotello},
  {Facchini}, {van der Marel}, {Testi}, \& {van
  Dishoeck}}]{2017AJ....153..240A}
{Ansdell}, M., {Williams}, J.~P., {Manara}, C.~F., {et~al.} 2017, \aj, 153,
  240, \dodoi{10.3847/1538-3881/aa69c0}

\bibitem[{{Azimlu} {et~al.}(2015){Azimlu}, {Mart{\'\i}nez-Galarza}, \&
  {Muench}}]{2015AJ....150...95A}
{Azimlu}, M., {Mart{\'\i}nez-Galarza}, J.~R., \& {Muench}, A.~A. 2015, \aj,
  150, 95, \dodoi{10.1088/0004-6256/150/3/95}

\bibitem[{{Barenfeld} {et~al.}(2016){Barenfeld}, {Carpenter}, {Ricci}, \&
  {Isella}}]{2016ApJ...827..142B}
{Barenfeld}, S.~A., {Carpenter}, J.~M., {Ricci}, L., \& {Isella}, A. 2016,
  \apj, 827, 142, \dodoi{10.3847/0004-637X/827/2/142}

\bibitem[{{Benisty} {et~al.}(2018){Benisty}, {Juh{\'a}sz}, {Facchini},
  {Pinilla}, {de Boer}, {P{\'e}rez}, {Keppler}, {Muro-Arena}, {Villenave},
  {Andrews}, {Dominik}, {Dullemond}, {Gallenne}, {Garufi}, {Ginski}, \&
  {Isella}}]{2018A&A...619A.171B}
{Benisty}, M., {Juh{\'a}sz}, A., {Facchini}, S., {et~al.} 2018, \aap, 619,
  A171, \dodoi{10.1051/0004-6361/201833913}

\bibitem[{{Birnstiel} {et~al.}(2010){Birnstiel}, {Dullemond}, \&
  {Brauer}}]{2010A&A...513A..79B}
{Birnstiel}, T., {Dullemond}, C.~P., \& {Brauer}, F. 2010, \aap, 513, A79,
  \dodoi{10.1051/0004-6361/200913731}

\bibitem[{{Bonnarel} {et~al.}(2000){Bonnarel}, {Fernique}, {Bienaym{\'e}},
  {Egret}, {Genova}, {Louys}, {Ochsenbein}, {Wenger}, \&
  {Bartlett}}]{2000A&AS..143...33B}
{Bonnarel}, F., {Fernique}, P., {Bienaym{\'e}}, O., {et~al.} 2000, \aaps, 143,
  33, \dodoi{10.1051/aas:2000331}

\bibitem[{{Bressan} {et~al.}(2012){Bressan}, {Marigo}, {Girardi}, {Salasnich},
  {Dal Cero}, {Rubele}, \& {Nanni}}]{2012MNRAS.427..127B}
{Bressan}, A., {Marigo}, P., {Girardi}, L., {et~al.} 2012, \mnras, 427, 127,
  \dodoi{10.1111/j.1365-2966.2012.21948.x}

\bibitem[{{Chabrier} \& {Baraffe}(1997)}]{1997A&A...327.1039C}
{Chabrier}, G., \& {Baraffe}, I. 1997, \aap, 327, 1039,
  \dodoi{10.48550/arXiv.astro-ph/9704118}

\bibitem[{{Chen} {et~al.}(2014){Chen}, {Girardi}, {Bressan}, {Marigo},
  {Barbieri}, \& {Kong}}]{2014MNRAS.444.2525C}
{Chen}, Y., {Girardi}, L., {Bressan}, A., {et~al.} 2014, \mnras, 444, 2525,
  \dodoi{10.1093/mnras/stu1605}

\bibitem[{{Choi} {et~al.}(2016){Choi}, {Dotter}, {Conroy}, {Cantiello},
  {Paxton}, \& {Johnson}}]{2016ApJ...823..102C}
{Choi}, J., {Dotter}, A., {Conroy}, C., {et~al.} 2016, \apj, 823, 102,
  \dodoi{10.3847/0004-637X/823/2/102}

\bibitem[{{Cifuentes} {et~al.}(2020){Cifuentes}, {Caballero},
  {Cort{\'e}s-Contreras}, {Montes}, {Abell{\'a}n}, {Dorda}, {Holgado},
  {Zapatero Osorio}, {Morales}, {Amado}, {Passegger}, {Quirrenbach}, {Reiners},
  {Ribas}, {Sanz-Forcada}, {Schweitzer}, {Seifert}, \&
  {Solano}}]{2020A&A...642A.115C}
{Cifuentes}, C., {Caballero}, J.~A., {Cort{\'e}s-Contreras}, M., {et~al.} 2020,
  \aap, 642, A115, \dodoi{10.1051/0004-6361/202038295}

\bibitem[{{Contreras Pe{\~n}a} {et~al.}(2019){Contreras Pe{\~n}a}, {Naylor}, \&
  {Morrell}}]{2019MNRAS.486.4590C}
{Contreras Pe{\~n}a}, C., {Naylor}, T., \& {Morrell}, S. 2019, \mnras, 486,
  4590, \dodoi{10.1093/mnras/stz1019}

\bibitem[{{Cutri} \& {et al.}(2014)}]{2014yCat.2328....0C}
{Cutri}, R.~M., \& {et al.} 2014, VizieR Online Data Catalog, II/328

\bibitem[{{Cutri} {et~al.}(2003){Cutri}, {Skrutskie}, {van Dyk}, {Beichman},
  {Carpenter}, {Chester}, {Cambresy}, {Evans}, {Fowler}, {Gizis}, {Howard},
  {Huchra}, {Jarrett}, {Kopan}, {Kirkpatrick}, {Light}, {Marsh}, {McCallon},
  {Schneider}, {Stiening}, {Sykes}, {Weinberg}, {Wheaton}, {Wheelock}, \&
  {Zacarias}}]{2003yCat.2246....0C}
{Cutri}, R.~M., {Skrutskie}, M.~F., {van Dyk}, S., {et~al.} 2003, VizieR Online
  Data Catalog, II/246

\bibitem[{{Dahm}(2008)}]{2008AJ....136..521D}
{Dahm}, S.~E. 2008, \aj, 136, 521, \dodoi{10.1088/0004-6256/136/2/521}

\bibitem[{{de Zeeuw} {et~al.}(1999){de Zeeuw}, {Hoogerwerf}, {de Bruijne},
  {Brown}, \& {Blaauw}}]{1999AJ....117..354D}
{de Zeeuw}, P.~T., {Hoogerwerf}, R., {de Bruijne}, J.~H.~J., {Brown}, A.~G.~A.,
  \& {Blaauw}, A. 1999, \aj, 117, 354, \dodoi{10.1086/300682}

\bibitem[{{Deng} {et~al.}(2012){Deng}, {Newberg}, {Liu}, {Carlin}, {Beers},
  {Chen}, {Chen}, {Christlieb}, {Grillmair}, {Guhathakurta}, {Han}, {Hou},
  {Lee}, {L{\'e}pine}, {Li}, {Liu}, {Pan}, {Sellwood}, {Wang}, {Wang}, {Yang},
  {Yanny}, {Zhang}, {Zhang}, {Zheng}, \& {Zhu}}]{2012RAA....12..735D}
{Deng}, L.-C., {Newberg}, H.~J., {Liu}, C., {et~al.} 2012, Research in
  Astronomy and Astrophysics, 12, 735, \dodoi{10.1088/1674-4527/12/7/003}

\bibitem[{{Du} {et~al.}(2021){Du}, {Luo}, {Zhang}, {Kong}, {Guo}, {Li}, {Zuo},
  {Wang}, {Chen}, \& {Zhao}}]{2021RAA....21..202D}
{Du}, B., {Luo}, A.~L., {Zhang}, S., {et~al.} 2021, Research in Astronomy and
  Astrophysics, 21, 202, \dodoi{10.1088/1674-4527/21/8/202}

\bibitem[{{Esplin} {et~al.}(2014){Esplin}, {Luhman}, \&
  {Mamajek}}]{2014ApJ...784..126E}
{Esplin}, T.~L., {Luhman}, K.~L., \& {Mamajek}, E.~E. 2014, \apj, 784, 126,
  \dodoi{10.1088/0004-637X/784/2/126}

\bibitem[{{Esplin} {et~al.}(2018){Esplin}, {Luhman}, {Miller}, \&
  {Mamajek}}]{2018AJ....156...75E}
{Esplin}, T.~L., {Luhman}, K.~L., {Miller}, E.~B., \& {Mamajek}, E.~E. 2018,
  \aj, 156, 75, \dodoi{10.3847/1538-3881/aacce0}

\bibitem[{{Fang} {et~al.}(2009){Fang}, {van Boekel}, {Wang}, {Carmona},
  {Sicilia-Aguilar}, \& {Henning}}]{2009AA...504..461F}
{Fang}, M., {van Boekel}, R., {Wang}, W., {et~al.} 2009, \aap, 504, 461,
  \dodoi{10.1051/0004-6361/200912468}

\bibitem[{{Fang} {et~al.}(2017){Fang}, {Kim}, {Pascucci}, {Apai}, {Zhang},
  {Sicilia-Aguilar}, {Alonso-Mart{\'\i}nez}, {Eiroa}, \&
  {Wang}}]{2017AJ....153..188F}
{Fang}, M., {Kim}, J.~S., {Pascucci}, I., {et~al.} 2017, \aj, 153, 188,
  \dodoi{10.3847/1538-3881/aa647b}

\bibitem[{{Farias} {et~al.}(2020){Farias}, {Tan}, \&
  {Eyer}}]{2020ApJ...900...14F}
{Farias}, J.~P., {Tan}, J.~C., \& {Eyer}, L. 2020, \apj, 900, 14,
  \dodoi{10.3847/1538-4357/aba699}

\bibitem[{{Francis} \& {van der Marel}(2020)}]{2020ApJ...892..111F}
{Francis}, L., \& {van der Marel}, N. 2020, \apj, 892, 111,
  \dodoi{10.3847/1538-4357/ab7b63}

\bibitem[{{Gaia Collaboration}(2020)}]{2020yCat.1350....0G}
{Gaia Collaboration}. 2020, VizieR Online Data Catalog, I/350

\bibitem[{{G{\'a}rate} {et~al.}(2021){G{\'a}rate}, {Delage}, {Stadler},
  {Pinilla}, {Birnstiel}, {Stammler}, {Picogna}, {Ercolano}, {Franz}, \&
  {Lenz}}]{2021A&A...655A..18G}
{G{\'a}rate}, M., {Delage}, T.~N., {Stadler}, J., {et~al.} 2021, \aap, 655,
  A18, \dodoi{10.1051/0004-6361/202141444}

\bibitem[{{Garaud} {et~al.}(2013){Garaud}, {Meru}, {Galvagni}, \&
  {Olczak}}]{2013ApJ...764..146G}
{Garaud}, P., {Meru}, F., {Galvagni}, M., \& {Olczak}, C. 2013, \apj, 764, 146,
  \dodoi{10.1088/0004-637X/764/2/146}

\bibitem[{{Hartigan} {et~al.}(1989){Hartigan}, {Hartmann}, {Kenyon}, {Hewett},
  \& {Stauffer}}]{1989ApJS...70..899H}
{Hartigan}, P., {Hartmann}, L., {Kenyon}, S., {Hewett}, R., \& {Stauffer}, J.
  1989, \apjs, 70, 899, \dodoi{10.1086/191361}

\bibitem[{{Hartigan} \& {Kenyon}(2003)}]{2003ApJ...583..334H}
{Hartigan}, P., \& {Kenyon}, S.~J. 2003, \apj, 583, 334, \dodoi{10.1086/345293}

\bibitem[{{Herbig}(1962)}]{1962AdA&A...1...47H}
{Herbig}, G.~H. 1962, Advances in Astronomy and Astrophysics, 1, 47,
  \dodoi{10.1016/B978-1-4831-9919-1.50006-6}

\bibitem[{{Herczeg} \& {Hillenbrand}(2014)}]{2014ApJ...786...97H}
{Herczeg}, G.~J., \& {Hillenbrand}, L.~A. 2014, \apj, 786, 97,
  \dodoi{10.1088/0004-637X/786/2/97}

\bibitem[{{Hoffer} \& {Ailon}(2014)}]{2014arXiv1412.6622H}
{Hoffer}, E., \& {Ailon}, N. 2014, arXiv e-prints, arXiv:1412.6622.
\newblock \doarXiv{1412.6622}

\bibitem[{{Hubickyj} {et~al.}(2005){Hubickyj}, {Bodenheimer}, \&
  {Lissauer}}]{2005Icar..179..415H}
{Hubickyj}, O., {Bodenheimer}, P., \& {Lissauer}, J.~J. 2005, \icarus, 179,
  415, \dodoi{10.1016/j.icarus.2005.06.021}

\bibitem[{{Husser} {et~al.}(2013){Husser}, {Wende-von Berg}, {Dreizler},
  {Homeier}, {Reiners}, {Barman}, \& {Hauschildt}}]{2013A&A...553A...6H}
{Husser}, T.~O., {Wende-von Berg}, S., {Dreizler}, S., {et~al.} 2013, \aap,
  553, A6, \dodoi{10.1051/0004-6361/201219058}

\bibitem[{{Ikoma} {et~al.}(2000){Ikoma}, {Nakazawa}, \&
  {Emori}}]{2000ApJ...537.1013I}
{Ikoma}, M., {Nakazawa}, K., \& {Emori}, H. 2000, \apj, 537, 1013,
  \dodoi{10.1086/309050}

\bibitem[{{Jao} {et~al.}(2018){Jao}, {Henry}, {Gies}, \&
  {Hambly}}]{2018ApJ...861L..11J}
{Jao}, W.-C., {Henry}, T.~J., {Gies}, D.~R., \& {Hambly}, N.~C. 2018, \apjl,
  861, L11, \dodoi{10.3847/2041-8213/aacdf6}

\bibitem[{{Kenyon} {et~al.}(2016){Kenyon}, {Najita}, \&
  {Bromley}}]{2016ApJ...831....8K}
{Kenyon}, S.~J., {Najita}, J.~R., \& {Bromley}, B.~C. 2016, \apj, 831, 8,
  \dodoi{10.3847/0004-637X/831/1/8}

\bibitem[{{Kessler-Silacci} {et~al.}(2006){Kessler-Silacci}, {Augereau},
  {Dullemond}, {Geers}, {Lahuis}, {Evans}, {van Dishoeck}, {Blake}, {Boogert},
  {Brown}, {J{\o}rgensen}, {Knez}, \& {Pontoppidan}}]{2006ApJ...639..275K}
{Kessler-Silacci}, J., {Augereau}, J.-C., {Dullemond}, C.~P., {et~al.} 2006,
  \apj, 639, 275, \dodoi{10.1086/499330}

\bibitem[{{Kirkpatrick} {et~al.}(1991){Kirkpatrick}, {Henry}, \&
  {McCarthy}}]{1991ApJS...77..417K}
{Kirkpatrick}, J.~D., {Henry}, T.~J., \& {McCarthy}, Donald~W., J. 1991, \apjs,
  77, 417, \dodoi{10.1086/191611}

\bibitem[{{Kirkpatrick} {et~al.}(2010){Kirkpatrick}, {Looper}, {Burgasser},
  {Schurr}, {Cutri}, {Cushing}, {Cruz}, {Sweet}, {Knapp}, {Barman},
  {Bochanski}, {Roellig}, {McLean}, {McGovern}, \&
  {Rice}}]{2010ApJS..190..100K}
{Kirkpatrick}, J.~D., {Looper}, D.~L., {Burgasser}, A.~J., {et~al.} 2010,
  \apjs, 190, 100, \dodoi{10.1088/0067-0049/190/1/100}

\bibitem[{{Komaki} {et~al.}(2021){Komaki}, {Nakatani}, \&
  {Yoshida}}]{2021ApJ...910...51K}
{Komaki}, A., {Nakatani}, R., \& {Yoshida}, N. 2021, \apj, 910, 51,
  \dodoi{10.3847/1538-4357/abe2af}

\bibitem[{{Kounkel} {et~al.}(2022){Kounkel}, {Stassun}, {Covey}, \&
  {Hartmann}}]{2022MNRAS.517..161K}
{Kounkel}, M., {Stassun}, K.~G., {Covey}, K., \& {Hartmann}, L. 2022, \mnras,
  517, 161, \dodoi{10.1093/mnras/stac2695}

\bibitem[{{Kraus} {et~al.}(2012){Kraus}, {Ireland}, {Hillenbrand}, \&
  {Martinache}}]{2012ApJ...745...19K}
{Kraus}, A.~L., {Ireland}, M.~J., {Hillenbrand}, L.~A., \& {Martinache}, F.
  2012, \apj, 745, 19, \dodoi{10.1088/0004-637X/745/1/19}

\bibitem[{{Kurosawa} {et~al.}(2006){Kurosawa}, {Harries}, \&
  {Symington}}]{2006MNRAS.370..580K}
{Kurosawa}, R., {Harries}, T.~J., \& {Symington}, N.~H. 2006, \mnras, 370, 580,
  \dodoi{10.1111/j.1365-2966.2006.10527.x}

\bibitem[{{Lambrechts} {et~al.}(2019){Lambrechts}, {Morbidelli}, {Jacobson},
  {Johansen}, {Bitsch}, {Izidoro}, \& {Raymond}}]{2019A&A...627A..83L}
{Lambrechts}, M., {Morbidelli}, A., {Jacobson}, S.~A., {et~al.} 2019, \aap,
  627, A83, \dodoi{10.1051/0004-6361/201834229}

\bibitem[{{Liu} {et~al.}(2019){Liu}, {Lambrechts}, {Johansen}, \&
  {Liu}}]{2019A&A...632A...7L}
{Liu}, B., {Lambrechts}, M., {Johansen}, A., \& {Liu}, F. 2019, \aap, 632, A7,
  \dodoi{10.1051/0004-6361/201936309}

\bibitem[{{Liu} {et~al.}(2020){Liu}, {Fu}, {Shi}, {Wu}, {Han}, {Chen}, {Dong},
  {Zhao}, {Chen}, {Zhang}, {Bai}, {Chen}, {Cui}, {Du}, {Hsia}, {Jiang}, {Hou},
  {Hou}, {Li}, {Li}, {Li}, {Liu}, {Liu}, {Luo}, {Ren}, {Tian}, {Tian}, {Wang},
  {Wu}, {Xie}, {Yan}, {Yang}, {Yu}, {Zhang}, {Zhang}, {Zhang}, {Zhang}, {Zhao},
  {Zhong}, {Zong}, \& {Zuo}}]{2020arXiv200507210L}
{Liu}, C., {Fu}, J., {Shi}, J., {et~al.} 2020, arXiv e-prints,
  arXiv:2005.07210, \dodoi{10.48550/arXiv.2005.07210}

\bibitem[{{Liu} {et~al.}(2014){Liu}, {Yuan}, {Huo}, {Deng}, {Hou}, {Zhao},
  {Zhao}, {Shi}, {Luo}, {Xiang}, {Zhang}, {Huang}, \&
  {Zhang}}]{2014IAUS..298..310L}
{Liu}, X.~W., {Yuan}, H.~B., {Huo}, Z.~Y., {et~al.} 2014, in Setting the scene
  for Gaia and LAMOST, ed. S.~{Feltzing}, G.~{Zhao}, N.~A. {Walton}, \&
  P.~{Whitelock}, Vol. 298, 310--321, \dodoi{10.1017/S1743921313006510}

\bibitem[{{Liu} {et~al.}(2022){Liu}, {Flock}, \& {Fang}}]{2022SCPMA..6569511L}
{Liu}, Y., {Flock}, M., \& {Fang}, M. 2022, Science China Physics, Mechanics,
  and Astronomy, 65, 269511, \dodoi{10.1007/s11433-021-1891-8}

\bibitem[{{Luhman}(2022{\natexlab{a}})}]{2022AJ....163...25L}
{Luhman}, K.~L. 2022{\natexlab{a}}, \aj, 163, 25,
  \dodoi{10.3847/1538-3881/ac35e3}

\bibitem[{{Luhman}(2022{\natexlab{b}})}]{2022AJ....163...24L}
---. 2022{\natexlab{b}}, \aj, 163, 24, \dodoi{10.3847/1538-3881/ac35e2}

\bibitem[{{Luhman}(2023)}]{2023AJ....165...37L}
---. 2023, \aj, 165, 37, \dodoi{10.3847/1538-3881/ac9da3}

\bibitem[{{Luhman} {et~al.}(2010){Luhman}, {Allen}, {Espaillat}, {Hartmann}, \&
  {Calvet}}]{2010ApJS..186..111L}
{Luhman}, K.~L., {Allen}, P.~R., {Espaillat}, C., {Hartmann}, L., \& {Calvet},
  N. 2010, \apjs, 186, 111, \dodoi{10.1088/0067-0049/186/1/111}

\bibitem[{{Luhman} \& {Mamajek}(2012)}]{2012ApJ...758...31L}
{Luhman}, K.~L., \& {Mamajek}, E.~E. 2012, \apj, 758, 31,
  \dodoi{10.1088/0004-637X/758/1/31}

\bibitem[{{Marton} {et~al.}(2019){Marton}, {{\'A}brah{\'a}m}, {Szegedi-Elek},
  {Varga}, {Kun}, {K{\'o}sp{\'a}l}, {Varga-Vereb{\'e}lyi}, {Hodgkin},
  {Szabados}, {Beck}, \& {Kiss}}]{2019MNRAS.487.2522M}
{Marton}, G., {{\'A}brah{\'a}m}, P., {Szegedi-Elek}, E., {et~al.} 2019, \mnras,
  487, 2522, \dodoi{10.1093/mnras/stz1301}

\bibitem[{{Meech} \& {Raymond}(2020)}]{2020plas.book..325M}
{Meech}, K., \& {Raymond}, S.~N. 2020, in Planetary Astrobiology, ed. V.~S.
  {Meadows}, G.~N. {Arney}, B.~E. {Schmidt}, \& D.~J. {Des Marais} (Tucson, AZ:
  University of Arizona Press), 325, \dodoi{10.2458/azu_uapress_9780816540068}

\bibitem[{{Meng} {et~al.}(2012){Meng}, {Rieke}, {Su}, {Ivanov}, {Vanzi}, \&
  {Rujopakarn}}]{2012ApJ...751L..17M}
{Meng}, H. Y.~A., {Rieke}, G.~H., {Su}, K. Y.~L., {et~al.} 2012, \apjl, 751,
  L17, \dodoi{10.1088/2041-8205/751/1/L17}

\bibitem[{{Miguel} {et~al.}(2020){Miguel}, {Cridland}, {Ormel}, {Fortney}, \&
  {Ida}}]{2020MNRAS.491.1998M}
{Miguel}, Y., {Cridland}, A., {Ormel}, C.~W., {Fortney}, J.~J., \& {Ida}, S.
  2020, \mnras, 491, 1998, \dodoi{10.1093/mnras/stz3007}

\bibitem[{{Montet} {et~al.}(2014){Montet}, {Crepp}, {Johnson}, {Howard}, \&
  {Marcy}}]{2014ApJ...781...28M}
{Montet}, B.~T., {Crepp}, J.~R., {Johnson}, J.~A., {Howard}, A.~W., \& {Marcy},
  G.~W. 2014, \apj, 781, 28, \dodoi{10.1088/0004-637X/781/1/28}

\bibitem[{{Morales} {et~al.}(2019){Morales}, {Mustill}, {Ribas}, {Davies},
  {Reiners}, {Bauer}, {Kossakowski}, {Herrero}, {Rodr{\'\i}guez},
  {L{\'o}pez-Gonz{\'a}lez}, {Rodr{\'\i}guez-L{\'o}pez}, {B{\'e}jar},
  {Gonz{\'a}lez-Cuesta}, {Luque}, {Pall{\'e}}, {Perger}, {Baroch}, {Johansen},
  {Klahr}, {Mordasini}, {Anglada-Escud{\'e}}, {Caballero},
  {Cort{\'e}s-Contreras}, {Dreizler}, {Lafarga}, {Nagel}, {Passegger},
  {Reffert}, {Rosich}, {Schweitzer}, {Tal-Or}, {Trifonov}, {Zechmeister},
  {Quirrenbach}, {Amado}, {Guenther}, {Hagen}, {Henning}, {Jeffers},
  {Kaminski}, {K{\"u}rster}, {Montes}, {Seifert}, {Abell{\'a}n}, {Abril},
  {Aceituno}, {Aceituno}, {Alonso-Floriano}, {Ammler-von Eiff}, {Antona},
  {Arroyo-Torres}, {Azzaro}, {Barrado}, {Becerril-Jarque}, {Ben{\'\i}tez},
  {Berdi{\~n}as}, {Bergond}, {Brinkm{\"o}ller}, {del Burgo}, {Burn},
  {Calvo-Ortega}, {Cano}, {C{\'a}rdenas}, {Cardona Guill{\'e}n}, {Carro},
  {Casal}, {Casanova}, {Casasayas-Barris}, {Chaturvedi}, {Cifuentes}, {Claret},
  {Colom{\'e}}, {Czesla}, {D{\'\i}ez-Alonso}, {Dorda}, {Emsenhuber},
  {Fern{\'a}ndez}, {Fern{\'a}ndez-Mart{\'\i}n}, {Ferro}, {Fuhrmeister},
  {Galad{\'\i}-Enr{\'\i}quez}, {Gallardo Cava}, {Garc{\'\i}a Vargas},
  {Garcia-Piquer}, {Gesa}, {Gonz{\'a}lez-{\'A}lvarez}, {Gonz{\'a}lez
  Hern{\'a}ndez}, {Gonz{\'a}lez-Peinado}, {Gu{\`a}rdia}, {Guijarro}, {de
  Guindos}, {Hatzes}, {Hauschildt}, {Hedrosa}, {Hermelo}, {Hern{\'a}ndez
  Arabi}, {Hern{\'a}ndez Otero}, {Hintz}, {Holgado}, {Huber}, {Huke},
  {Johnson}, {de Juan}, {Kehr}, {Kemmer}, {Kim}, {Kl{\"u}ter}, {Klutsch},
  {Labarga}, {Labiche}, {Lalitha}, {Lamp{\'o}n}, {Lara}, {Launhardt},
  {L{\'a}zaro}, {Lizon}, {Llamas}, {Lodieu}, {L{\'o}pez del Fresno}, {L{\'o}pez
  Salas}, {L{\'o}pez-Santiago}, {Mag{\'a}n Madinabeitia}, {Mall}, {Mancini},
  {Mandel}, {Marfil}, {Mar{\'\i}n Molina}, {Mart{\'\i}n},
  {Mart{\'\i}n-Fern{\'a}ndez}, {Mart{\'\i}n-Ruiz},
  {Mart{\'\i}nez-Rodr{\'\i}guez}, {Marvin}, {Mirabet}, {Moya}, {Naranjo},
  {Nelson}, {Nortmann}, {Nowak}, {Ofir}, {Pascual}, {Pavlov}, {Pedraz},
  {P{\'e}rez Medialdea}, {P{\'e}rez-Calpena}, {Perryman}, {Rabaza}, {Ram{\'o}n
  Ballesta}, {Rebolo}, {Redondo}, {Rix}, {Rodler}, {Rodr{\'\i}guez Trinidad},
  {Sabotta}, {Sadegi}, {Salz}, {S{\'a}nchez-Blanco}, {S{\'a}nchez Carrasco},
  {S{\'a}nchez-L{\'o}pez}, {Sanz-Forcada}, {Sarkis}, {Sarmiento},
  {Sch{\"a}fer}, {Schlecker}, {Schmitt}, {Sch{\"o}fer}, {Solano}, {Sota},
  {Stahl}, {Stock}, {Stuber}, {St{\"u}rmer}, {Su{\'a}rez}, {Tabernero},
  {Tulloch}, {Veredas}, {Vico-Linares}, {Vilardell}, {Wagner}, {Winkler},
  {Wolthoff}, {Yan}, \& {Zapatero Osorio}}]{2019Sci...365.1441M}
{Morales}, J.~C., {Mustill}, A.~J., {Ribas}, I., {et~al.} 2019, Science, 365,
  1441, \dodoi{10.1126/science.aax3198}

\bibitem[{{Morbidelli} {et~al.}(2012){Morbidelli}, {Lunine}, {O'Brien},
  {Raymond}, \& {Walsh}}]{2012AREPS..40..251M}
{Morbidelli}, A., {Lunine}, J.~I., {O'Brien}, D.~P., {Raymond}, S.~N., \&
  {Walsh}, K.~J. 2012, Annual Review of Earth and Planetary Sciences, 40, 251,
  \dodoi{10.1146/annurev-earth-042711-105319}

\bibitem[{{Mulders} {et~al.}(2021){Mulders}, {Dr{\k{a}}{\.z}kowska}, {van der
  Marel}, {Ciesla}, \& {Pascucci}}]{2021ApJ...920L...1M}
{Mulders}, G.~D., {Dr{\k{a}}{\.z}kowska}, J., {van der Marel}, N., {Ciesla},
  F.~J., \& {Pascucci}, I. 2021, \apjl, 920, L1,
  \dodoi{10.3847/2041-8213/ac2947}

\bibitem[{{M{\"u}ller} {et~al.}(2018){M{\"u}ller}, {Keppler}, {Henning},
  {Samland}, {Chauvin}, {Beust}, {Maire}, {Molaverdikhani}, {van Boekel},
  {Benisty}, {Boccaletti}, {Bonnefoy}, {Cantalloube}, {Charnay}, {Baudino},
  {Gennaro}, {Long}, {Cheetham}, {Desidera}, {Feldt}, {Fusco}, {Girard},
  {Gratton}, {Hagelberg}, {Janson}, {Lagrange}, {Langlois}, {Lazzoni}, {Ligi},
  {M{\'e}nard}, {Mesa}, {Meyer}, {Molli{\`e}re}, {Mordasini}, {Moulin},
  {Pavlov}, {Pawellek}, {Quanz}, {Ramos}, {Rouan}, {Sissa}, {Stadler}, {Vigan},
  {Wahhaj}, {Weber}, \& {Zurlo}}]{2018A&A...617L...2M}
{M{\"u}ller}, A., {Keppler}, M., {Henning}, T., {et~al.} 2018, \aap, 617, L2,
  \dodoi{10.1051/0004-6361/201833584}

\bibitem[{{Pecaut} \& {Mamajek}(2016)}]{2016MNRAS.461..794P}
{Pecaut}, M.~J., \& {Mamajek}, E.~E. 2016, \mnras, 461, 794,
  \dodoi{10.1093/mnras/stw1300}

\bibitem[{{P{\'e}rez} {et~al.}(2018){P{\'e}rez}, {Benisty}, {Andrews},
  {Isella}, {Dullemond}, {Huang}, {Kurtovic}, {Guzm{\'a}n}, {Zhu}, {Birnstiel},
  {Zhang}, {Carpenter}, {Wilner}, {Ricci}, {Bai}, {Weaver}, \&
  {{\"O}berg}}]{2018ApJ...869L..50P}
{P{\'e}rez}, L.~M., {Benisty}, M., {Andrews}, S.~M., {et~al.} 2018, \apjl, 869,
  L50, \dodoi{10.3847/2041-8213/aaf745}

\bibitem[{{Pfalzner} {et~al.}(2022){Pfalzner}, {Dehghani}, \&
  {Michel}}]{2022ApJ...939L..10P}
{Pfalzner}, S., {Dehghani}, S., \& {Michel}, A. 2022, \apjl, 939, L10,
  \dodoi{10.3847/2041-8213/ac9839}

\bibitem[{{Picogna} {et~al.}(2021){Picogna}, {Ercolano}, \&
  {Espaillat}}]{2021MNRAS.508.3611P}
{Picogna}, G., {Ercolano}, B., \& {Espaillat}, C.~C. 2021, \mnras, 508, 3611,
  \dodoi{10.1093/mnras/stab2883}

\bibitem[{{Pollack} {et~al.}(1996){Pollack}, {Hubickyj}, {Bodenheimer},
  {Lissauer}, {Podolak}, \& {Greenzweig}}]{1996Icar..124...62P}
{Pollack}, J.~B., {Hubickyj}, O., {Bodenheimer}, P., {et~al.} 1996, \icarus,
  124, 62, \dodoi{10.1006/icar.1996.0190}

\bibitem[{{Quirrenbach} {et~al.}(2022){Quirrenbach}, {Passegger}, {Trifonov},
  {Amado}, {Caballero}, {Reiners}, {Ribas}, {Aceituno}, {B{\'e}jar},
  {Chaturvedi}, {Gonz{\'a}lez-Cuesta}, {Henning}, {Herrero}, {Kaminski},
  {K{\"u}rster}, {Lalitha}, {Lodieu}, {L{\'o}pez-Gonz{\'a}lez}, {Montes},
  {Pall{\'e}}, {Perger}, {Pollacco}, {Reffert}, {Rodr{\'\i}guez}, {L{\'o}pez},
  {Shan}, {Tal-Or}, {Osorio}, \& {Zechmeister}}]{2022A&A...663A..48Q}
{Quirrenbach}, A., {Passegger}, V.~M., {Trifonov}, T., {et~al.} 2022, \aap,
  663, A48, \dodoi{10.1051/0004-6361/202142915}

\bibitem[{{Raymond} \& {Morbidelli}(2022)}]{2022ASSL..466....3R}
{Raymond}, S.~N., \& {Morbidelli}, A. 2022, in Astrophysics and Space Science
  Library, Vol. 466, Demographics of Exoplanetary Systems, Lecture Notes of the
  3rd Advanced School on Exoplanetary Science, ed. K.~{Biazzo}, V.~{Bozza},
  L.~{Mancini}, \& A.~{Sozzetti}, 3--82, \dodoi{10.1007/978-3-030-88124-5_1}

\bibitem[{{Reipurth} {et~al.}(1996){Reipurth}, {Pedrosa}, \&
  {Lago}}]{1996A&AS..120..229R}
{Reipurth}, B., {Pedrosa}, A., \& {Lago}, M.~T.~V.~T. 1996, \aaps, 120, 229

\bibitem[{{Riaud} {et~al.}(2006){Riaud}, {Mawet}, {Absil}, {Boccaletti},
  {Baudoz}, {Herwats}, \& {Surdej}}]{2006A&A...458..317R}
{Riaud}, P., {Mawet}, D., {Absil}, O., {et~al.} 2006, \aap, 458, 317,
  \dodoi{10.1051/0004-6361:20065232}

\bibitem[{{Ribas} {et~al.}(2014){Ribas}, {Mer{\'\i}n}, {Bouy}, \&
  {Maud}}]{2014A&A...561A..54R}
{Ribas}, {\'A}., {Mer{\'\i}n}, B., {Bouy}, H., \& {Maud}, L.~T. 2014, \aap,
  561, A54, \dodoi{10.1051/0004-6361/201322597}

\bibitem[{{Ruder}(2016)}]{2016arXiv160904747R}
{Ruder}, S. 2016, arXiv e-prints, arXiv:1609.04747.
\newblock \doarXiv{1609.04747}

\bibitem[{{Schlafly} {et~al.}(2014){Schlafly}, {Green}, {Finkbeiner}, {Rix},
  {Bell}, {Burgett}, {Chambers}, {Draper}, {Hodapp}, {Kaiser}, {Magnier},
  {Martin}, {Metcalfe}, {Price}, \& {Tonry}}]{2014ApJ...786...29S}
{Schlafly}, E.~F., {Green}, G., {Finkbeiner}, D.~P., {et~al.} 2014, \apj, 786,
  29, \dodoi{10.1088/0004-637X/786/1/29}

\bibitem[{{Schlecker} {et~al.}(2022){Schlecker}, {Burn}, {Sabotta}, {Seifert},
  {Henning}, {Emsenhuber}, {Mordasini}, {Reffert}, {Shan}, \&
  {Klahr}}]{2022A&A...664A.180S}
{Schlecker}, M., {Burn}, R., {Sabotta}, S., {et~al.} 2022, \aap, 664, A180,
  \dodoi{10.1051/0004-6361/202142543}

\bibitem[{{Schoettler} {et~al.}(2020){Schoettler}, {de Bruijne}, {Vaher}, \&
  {Parker}}]{2020MNRAS.495.3104S}
{Schoettler}, C., {de Bruijne}, J., {Vaher}, E., \& {Parker}, R.~J. 2020,
  \mnras, 495, 3104, \dodoi{10.1093/mnras/staa1228}

\bibitem[{{Sicilia-Aguilar} {et~al.}(2004){Sicilia-Aguilar}, {Hartmann},
  {Brice{\~n}o}, {Muzerolle}, \& {Calvet}}]{2004AJ....128..805S}
{Sicilia-Aguilar}, A., {Hartmann}, L.~W., {Brice{\~n}o}, C., {Muzerolle}, J.,
  \& {Calvet}, N. 2004, \aj, 128, 805, \dodoi{10.1086/422432}

\bibitem[{{Silverberg} {et~al.}(2020){Silverberg}, {Wisniewski}, {Kuchner},
  {Lawson}, {Bans}, {Debes}, {Biggs}, {Bosch}, {Doll}, {Luca}, {Enachioaie},
  {Hamilton}, {Holden}, \& {Hyogo}}]{2020ApJ...890..106S}
{Silverberg}, S.~M., {Wisniewski}, J.~P., {Kuchner}, M.~J., {et~al.} 2020,
  \apj, 890, 106, \dodoi{10.3847/1538-4357/ab68e6}

\bibitem[{{Skrutskie} {et~al.}(2006){Skrutskie}, {Cutri}, {Stiening},
  {Weinberg}, {Schneider}, {Carpenter}, {Beichman}, {Capps}, {Chester},
  {Elias}, {Huchra}, {Liebert}, {Lonsdale}, {Monet}, {Price}, {Seitzer},
  {Jarrett}, {Kirkpatrick}, {Gizis}, {Howard}, {Evans}, {Fowler}, {Fullmer},
  {Hurt}, {Light}, {Kopan}, {Marsh}, {McCallon}, {Tam}, {Van Dyk}, \&
  {Wheelock}}]{2006AJ....131.1163S}
{Skrutskie}, M.~F., {Cutri}, R.~M., {Stiening}, R., {et~al.} 2006, \aj, 131,
  1163, \dodoi{10.1086/498708}

\bibitem[{{Sousa} {et~al.}(2019){Sousa}, {Alencar}, {Rebull}, {Espaillat},
  {Calvet}, \& {Teixeira}}]{2019A&A...629A..67S}
{Sousa}, A.~P., {Alencar}, S. H.~P., {Rebull}, L.~M., {et~al.} 2019, \aap, 629,
  A67, \dodoi{10.1051/0004-6361/201935563}

\bibitem[{{Sun} {et~al.}(2021){Sun}, {Jiang}, {Zhao}, \&
  {Ren}}]{2021ApJS..256...46S}
{Sun}, M., {Jiang}, B., {Zhao}, H., \& {Ren}, Y. 2021, \apjs, 256, 46,
  \dodoi{10.3847/1538-4365/ac1601}

\bibitem[{{Tajiri} {et~al.}(2020){Tajiri}, {Kawahara}, {Aizawa}, {Fujii},
  {Hattori}, {Kasagi}, {Kotani}, {Masuda}, {Momose}, {Muto}, {Ohsawa}, \&
  {Takita}}]{2020ApJS..251...18T}
{Tajiri}, T., {Kawahara}, H., {Aizawa}, M., {et~al.} 2020, \apjs, 251, 18,
  \dodoi{10.3847/1538-4365/abbc17}

\bibitem[{{Tang} {et~al.}(2014){Tang}, {Bressan}, {Rosenfield}, {Slemer},
  {Marigo}, {Girardi}, \& {Bianchi}}]{2014MNRAS.445.4287T}
{Tang}, J., {Bressan}, A., {Rosenfield}, P., {et~al.} 2014, \mnras, 445, 4287,
  \dodoi{10.1093/mnras/stu2029}

\bibitem[{{Thanathibodee} {et~al.}(2022){Thanathibodee}, {Calvet},
  {Hern{\'a}ndez}, {Mauc{\'o}}, \& {Brice{\~n}o}}]{2022AJ....163...74T}
{Thanathibodee}, T., {Calvet}, N., {Hern{\'a}ndez}, J., {Mauc{\'o}}, K., \&
  {Brice{\~n}o}, C. 2022, \aj, 163, 74, \dodoi{10.3847/1538-3881/ac3ee6}

\bibitem[{{Wagner} {et~al.}(2018){Wagner}, {Follete}, {Close}, {Apai}, {Gibbs},
  {Keppler}, {M{\"u}ller}, {Henning}, {Kasper}, {Wu}, {Long}, {Males},
  {Morzinski}, \& {McClure}}]{2018ApJ...863L...8W}
{Wagner}, K., {Follete}, K.~B., {Close}, L.~M., {et~al.} 2018, \apjl, 863, L8,
  \dodoi{10.3847/2041-8213/aad695}

\bibitem[{{Wang} \& {Chen}(2019)}]{2019ApJ...877..116W}
{Wang}, S., \& {Chen}, X. 2019, \apj, 877, 116,
  \dodoi{10.3847/1538-4357/ab1c61}

\bibitem[{{Wells} {et~al.}(2021){Wells}, {Rackham}, {Schanche}, {Petrucci},
  {G{\'o}mez Maqueo Chew}, {Demory}, {Burgasser}, {Burn}, {Pozuelos},
  {G{\"u}nther}, {Sabin}, {Schroffenegger}, {G{\'o}mez-Mu{\~n}oz}, {Stassun},
  {Van Grootel}, {Howell}, {Sebastian}, {Triaud}, {Apai}, {Plauchu-Frayn},
  {Guerrero}, {Guill{\'e}n}, {Landa}, {Melgoza}, {Montalvo}, {Serrano},
  {Riesgo}, {Barkaoui}, {Bixel}, {Burdanov}, {Chen}, {Chinchilla}, {Collins},
  {Daylan}, {de Wit}, {Delrez}, {D{\'e}vora-Pajares}, {Dietrich}, {Dransfield},
  {Ducrot}, {Fausnaugh}, {Furlan}, {Gabor}, {Gan}, {Garcia}, {Ghachoui},
  {Giacalone}, {Gibbs}, {Gillon}, {Gnilka}, {Gore}, {Guerrero}, {Henning},
  {Hesse}, {Jehin}, {Jenkins}, {Latham}, {Lester}, {McCormac}, {Murray},
  {Niraula}, {Pedersen}, {Queloz}, {Ricker}, {Rodriguez}, {Schroeder},
  {Schwarz}, {Scott}, {Seager}, {Theissen}, {Thompson}, {Timmermans},
  {Twicken}, \& {Winn}}]{2021A&A...653A..97W}
{Wells}, R.~D., {Rackham}, B.~V., {Schanche}, N., {et~al.} 2021, \aap, 653,
  A97, \dodoi{10.1051/0004-6361/202141277}

\bibitem[{{Wenger} {et~al.}(2000){Wenger}, {Ochsenbein}, {Egret}, {Dubois},
  {Bonnarel}, {Borde}, {Genova}, {Jasniewicz}, {Lalo{\"e}}, {Lesteven}, \&
  {Monier}}]{2000A&AS..143....9W}
{Wenger}, M., {Ochsenbein}, F., {Egret}, D., {et~al.} 2000, \aaps, 143, 9,
  \dodoi{10.1051/aas:2000332}

\bibitem[{{White} \& {Basri}(2003)}]{2003ApJ...582.1109W}
{White}, R.~J., \& {Basri}, G. 2003, \apj, 582, 1109, \dodoi{10.1086/344673}

\bibitem[{{Xiang} {et~al.}(2017){Xiang}, {Liu}, {Yuan}, {Huo}, {Huang}, {Wang},
  {Chen}, {Ren}, {Zhang}, {Tian}, {Yang}, {Shi}, {Zhao}, {Li}, {Zhao}, {Cui},
  {Li}, {Hou}, {Zhang}, {Zhang}, {Wang}, {Wu}, {Cao}, {Yan}, {Yan}, {Luo},
  {Zhang}, {Bai}, {Yuan}, {Dong}, {Lei}, \& {Li}}]{2017MNRAS.467.1890X}
{Xiang}, M.~S., {Liu}, X.~W., {Yuan}, H.~B., {et~al.} 2017, \mnras, 467, 1890,
  \dodoi{10.1093/mnras/stx129}

\bibitem[{{Yao} {et~al.}(2018){Yao}, {Meyer}, {Covey}, {Tan}, \& {Da
  Rio}}]{2018ApJ...869...72Y}
{Yao}, Y., {Meyer}, M.~R., {Covey}, K.~R., {Tan}, J.~C., \& {Da Rio}, N. 2018,
  \apj, 869, 72, \dodoi{10.3847/1538-4357/aaec7a}

\bibitem[{{Yuan} {et~al.}(2015){Yuan}, {Liu}, {Huo}, {Xiang}, {Huang}, {Chen},
  {Zhang}, {Sun}, {Wang}, {Zhang}, {Zhao}, {Luo}, {Shi}, {Li}, {Yuan}, {Dong},
  {Li}, {Hou}, \& {Zhang}}]{2015MNRAS.448..855Y}
{Yuan}, H.~B., {Liu}, X.~W., {Huo}, Z.~Y., {et~al.} 2015, \mnras, 448, 855,
  \dodoi{10.1093/mnras/stu2723}

\bibitem[{{Zhao} {et~al.}(2012){Zhao}, {Zhao}, {Chu}, {Jing}, \&
  {Deng}}]{2012RAA....12..723Z}
{Zhao}, G., {Zhao}, Y.-H., {Chu}, Y.-Q., {Jing}, Y.-P., \& {Deng}, L.-C. 2012,
  Research in Astronomy and Astrophysics, 12, 723,
  \dodoi{10.1088/1674-4527/12/7/002}

\bibitem[{{Zucker} {et~al.}(2020){Zucker}, {Speagle}, {Schlafly}, {Green},
  {Finkbeiner}, {Goodman}, \& {Alves}}]{2020A&A...633A..51Z}
{Zucker}, C., {Speagle}, J.~S., {Schlafly}, E.~F., {et~al.} 2020, \aap, 633,
  A51, \dodoi{10.1051/0004-6361/201936145}

\bibitem[{{Zucker} {et~al.}(2019){Zucker}, {Speagle}, {Schlafly}, {Green},
  {Finkbeiner}, {Goodman}, \& {Alves}}]{2019ApJ...879..125Z}
---. 2019, \apj, 879, 125, \dodoi{10.3847/1538-4357/ab2388}

\end{thebibliography}
\bibliographystyle{aasjournal}

 \end{CJK*}
\end{document}